\documentclass[%
superscriptaddress,
twocolumn,
 amsmath,amssymb,
aps,
]{revtex4-1} 
\usepackage[english]{babel} 
\usepackage[T1]{fontenc}
\usepackage{selinput}
\SelectInputMappings{%
  aacute={á},
  ntilde={ñ},
  Euro={€}
}
\usepackage{braket}
\usepackage{babel}
\usepackage{graphicx}
\usepackage{dcolumn}
\usepackage{bm}

\usepackage{soul}
 
\usepackage{hyperref}
\hypersetup{hidelinks,
backref=true,
pagebackref=true,
hyperindex=true,
breaklinks=true,
colorlinks=true,linkcolor=black,citecolor=black,
urlcolor=blue,
bookmarks=true,
bookmarksopen=false,
pdftitle={Title},
pdfauthor={Author}}
\newcommand{\ipcms}{\mbox{Universit\'e de Strasbourg, CNRS, Institut de Physique et Chimie des Mat\'eriaux 
de Strasbourg,} UMR 7504, F-67000 Strasbourg, France}
\newcommand{\uba}{Departamento de F\'isica ``J. J. Giambiagi'' and IFIBA, FCEyN, Universidad de Buenos Aires, 1428 Buenos Aires, Argentina}
\newcommand{\ifimar}{Instituto de Investigaciones F\'isicas de Mar del Plata (IFIMAR), Facultad de Ciencias Exactas y Naturales, Universidad Nacional de Mar del Plata, CONICET, 7600 Mar del Plata, Argentina}

\begin{document}
\title{Gauging classical and quantum integrability through out-of-time-ordered correlators}

\author{Emiliano M. Fortes}
\affiliation{\uba}%
\author{Ignacio  Garc\'ia-Mata}
\affiliation{\ifimar} 
\author{Rodolfo  A.  Jalabert}
\affiliation{\ipcms}%
\author{Diego A. Wisniacki}
\affiliation{\uba}

\date{\today}

\begin{abstract}
Out-of-time-ordered correlators (OTOCs) have been proposed as a probe of chaos in quantum mechanics, on the basis of their short-time exponential growth found in some particular set-ups. However, it has been seen that this behavior is not universal. Therefore, we query other quantum chaos manifestations arising from the OTOCs and we thus study their long-time behavior in systems of completely different nature: quantum maps, which are the simplest chaotic one-body system and spin chains, which are many-body systems without a classical limit. It is shown that studying the long-time regime of the OTOCs it is possible to detect and gauge the 
transition between integrability and chaos, and we benchmark the transition with other indicators of quantum chaos based on the spectra and the eigenstates of the systems considered. For systems with a classical analog, we show that the proposed OTOC indicators have a very high accuracy 
that allow us to detect subtle features along the integrability-to-chaos transition. 
\end{abstract}
\maketitle

\section{\label{sec:level1}Introduction}
The original Bohr-Sommerfeld formulation of quantum mechanics addressed integrable classical systems, with as many conserved quantities as degrees of freedom. Einstein’s 1917 observation that such a quantization scheme remained extremely limited (as integrability is a singularity among dynamical systems) remained relatively unnoticed until the late 1950s, probably due to the success of the Schr\"odinger equation \cite{einstein,stone}. The quantization of chaotic systems, as well as the understanding of the consequences of classical chaos on quantum observables such as the level statistics, developed in the 1970s and 1980s \cite{gutzwiller,Bohigas}, provided the connection of quantum mechanics with fully chaotic systems, which constitute another singularity within the ensemble of dynamical systems. The connection of classical and quantum properties in the generic case of mixed systems, away from the two previously mentioned singularities, remains, comparatively, less understood and more difficult to quantify. 

Two important aspects spur our interest in the intermediate behavior between fully integrable and completely chaotic regimes. On the one hand, in quantum systems without a classical analog the notion of integrability is still valid, although defined through the separability and the soluble character of the quantum problem. On the other hand, in multidimensional and many-body systems the ``no-man's land'' between the two singular behaviors is difficult to avoid. 

Within this context, the recent impressive development of experimental techniques for many-body quantum systems \cite{LewisSwan:2019gj,Monroe1,Kaufman794,Schreiber842,Li:2017ic}, monitoring in time complex processes like localization or thermalization, enhanced the need to understand quantum dynamics and its connection with the concepts of integrability, chaos and ergodicity. A useful tool towards this task, which has lately received considerable attention is the out-of-time ordered correlator (OTOC), defined from the commutator of two operators $\hat{V}$ and $\hat{W}(t)$ 
(the Heisenberg time evolution of operator $\hat{W})$ as
\begin{equation}
C(t) = \left\langle [\hat{W}(t),\hat{V}]^\dagger[\hat{W}(t),\hat{V}]\right\rangle \, ,
\label{eq:OTOC1}
\end{equation}
where  the angular brackets denote the average over an initial state. While this time-dependent quantity was first considered in a semi-classical study of superconductivity \cite{1969JETP...28.1200L}, the present interest results from its use as a measure of quantum information scrambling \cite{Bound,PhysRevA.94.040302,Riddell:2019ed,Landsman:2019ke,SwingleSlow,chen2017out,SlaglePRB2017,Luitz2017,Sahu}, which in addition, is accessible to experiments \cite{Li:2017ic,garttner2017measuring,cappellaro2018,pastawski2019}. 

For chaotic many-body systems the scrambling measured by the OTOC was conjectured to increase exponentially in time, with a temperature-dependent bound on the growth-rate \cite{Bound}. The strongly-coupled, exactly-solvable Sachdev-Ye-Kitaev many-body quantum model \cite{SachdevYe,Kitaev}, saturates the bound, while being dually related to black holes via the anti-de Sitter?conformal field theory (AdS-CFT) correspondence \cite{AdS}. 

The exponential short-time behavior has been demonstrated to hold in some many-body systems such as the Dicke \cite{ChavezCarlos:2019bn} and the 
Sachdev-Ye-Kitaev  \cite{Roberts:2018ck,Maldacena:2016hq} models, and the corresponding systems have then been dubbed  "fast scramblers" \cite{Sekino_2008}. Systems with a classical counterpart, like quantum maps and billiards, also exhibit an exponential growth of the OTOC, with a rate that can be either equal \cite{GarciaMata:2018kz,Jalabert:2018da} or proportional \cite{Rozenbaum2017,chen2018operator,lakshminarayan2019out} to the Lyapunov exponent, depending on the initial state. However, the exponential growth does not constitute a universal behavior. Other examples, like spin chains in the presence of random fields \cite{Riddell:2019ed}, Luttinger liquids \cite{Dora:2017go}, or weakly chaotic systems \cite{ProsenWeak2017} exhibit a polynomial increase of the OTOC (and have then been dubbed ``slow scramblers''). 

While the exponential growth of the OTOC for relatively short times was the initial focus for the above-cited studies, it later appeared that the long-time properties of the OTOC were equally interesting from a quantum chaos point of view \cite{rammensee2018many,GarciaMata:2018kz}. 

The comparatively fewer studies of the long-time behavior of the OTOC \cite{chen2018operator,rammensee2018many,GarciaMata:2018kz,Jalabert:2018da,hashimoto,cotler} have been centered on the saturation obtained in finite-size classically chaotic systems. A semi-classical theory for fully chaotic many-body systems linked the saturation with quantum interference  \cite{rammensee2018many}. In the simpler case of quantum maps, the approach to the saturation was shown to be exponential, and dominated by the largest nontrivial Ruelle-Pollicott resonance of their chaotic classical counterparts \cite{GarciaMata:2018kz}. More recently the long-time regime in the case of two interacting maps was also considered \cite{prakash2019scrambling,Bergamasco2019}.  In addition, the appearance of oscillations in the long-time regime for critical many-body systems was studied in Ref.\cite{hummel2018reversible}. The long-time behavior of the OTOC is particularly interesting in view of its consequences for thermalization processes in many-body systems  \cite{chan2018eigenstate,khemani2018velocity,borgonovi2019emergence,LewisSwan:2019gj,Bohrdt_2017}.

The goal of this work is to try to provide a connection between the degree of integrability and the characteristic features of the OTOC dynamics for the case of mixed systems. We will show that such a connection is firmly established using the long-time dynamics of the OTOC, where we can match the quantum and classical chaos indicator with proposed OTOC indicators. 

In order to test the universality of the established connection, we study one-particle systems having a classical counterpart (two quantum maps) where a parameter can be used to tune the transition from integrability to chaos, as well as many-particle systems without a classical analog (three different spin chains) where chaos is typically driven by an interaction parameter and characterized by the nearest-level spacing-distribution. 

This work has the following structure. 
In Sec.~\ref{sec:OTOC} we review the OTOC and its most  relevant properties  while in Sec.~\ref{sec:chaos_in} two of the canonical quantum  chaos indicators are exposed. 
Both previous sections provide the mechanisms to analyze the systems studied in this work. In Sec.~\ref{sec:sysres} we present the physical systems to be used in numerical simulations,  
consisting of quantum maps (Sec.~\ref{sec:qmaps}) and three many-body Spin chain systems (Sec.~\ref{sec:SCN}). We present our conclusions and outlook in 
 Sec.~\ref{sec:conclusions}. In  Appendix~\ref{ap:short_time} the mathematical details of the calculations for the short-time behavior in spin chains are included
 and in Appppendix~\ref{app:metro}
we explain a method to measure the area of the chaotic region in the phase space.
\section{\label{sec:OTOC} Short and long-time behavior of the out-of-time-ordered correlator}
We will work with initial thermal states in the infinite temperature limit, for which $\small{\left\langle \hat{O}\right\rangle = Tr\left\{ \hat{O}/D\right\}}$, where $D$ is the dimension of the Hilbert space. Moreover, taking the $\hat{W}$ and $\hat{V}$ operators to be Hermitian, the OTOC defined in Eq.~\eqref{eq:OTOC1} becomes
\begin{equation}
    C(t)=\small{\frac{2}{D}\left\{Tr(\hat{V} \hat{W}^2(t)\hat{V})-Re \left[Tr(\hat{W}(t)\hat{V}\hat{W}(t)\hat{V})\right] 
    \right\}} \, .
\label{eq:OTOCinftemp}
\end{equation}

\begin{figure}[h]
\centering
\includegraphics[width=.99\linewidth]{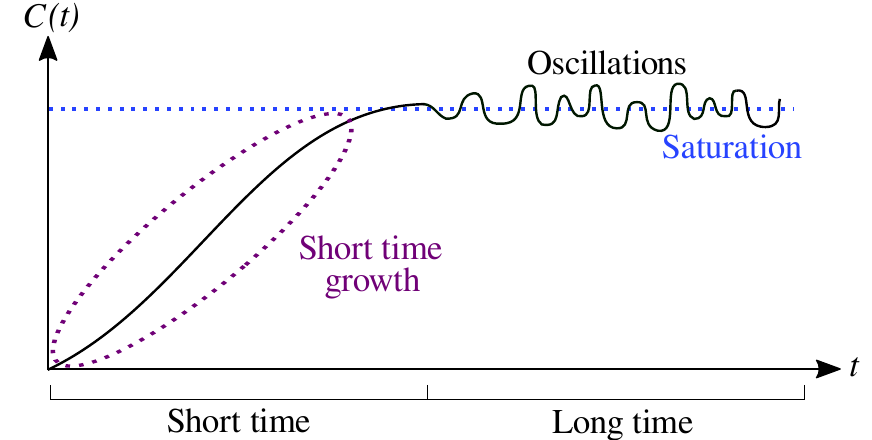}
\caption{Sketch of the typical time-dependence of the OTOC for one-body and finite-size many-body systems exhibiting different behavior in the short and long-time regimes.}
\label{Fig:OTOC_show}
\end{figure}

Typically, the time-dependence of the OTOC appears as schematically presented in Fig.~\ref{Fig:OTOC_show}, with two well-defined time regimes.
The short-time growth of the OTOC is given by the operator spread, or scrambling, where the initial quantum information spreads over the available degrees of freedom in a quantum system.
As discussed in the introduction, the short-time growth is exponential in many cases \cite{Bound,rammensee2018many,GarciaMata:2018kz,Rozenbaum2017,lakshminarayan2019out,ChavezCarlos:2019bn}. But such a behavior of fast scramblers is not generic. Examples of slow scramblers have been predicted for weakly chaotic systems, where the short-time growth has been shown to be linear \cite{ProsenWeak2017}, for Luttinger liquids, where a quadratic  initial-state-independent behavior has been obtained \cite{Dora:2017go}, for a random-field $XX$ spin chain, where the initial growth exhibits a power law given twice the distance between the sites associated with the chosen operators \cite{Riddell:2019ed}, as well as for the anomalous phases of the interacting Aubry-Andr\'e model \cite{dassarma}. The rich, nonuniversal initial behavior of the OTOC hinted at the usefulness of using it in order to characterize different many-body phases \cite{SwingleSlow,chen2017out,SlaglePRB2017,Luitz2017,xu2018accessing,xu2018locality,Sahu,dassarma}. 

The dotted region defined in violet in the sketch of Fig.~\ref{Fig:OTOC_show} stands for the variety of possible outcomes for the OTOC's short-time growth. Even if in this work we concentrate ourselves in the infinite-temperature limit, it is worth mentioning that in the general case of an initial thermal state, the extent of the growth regime can be strongly dependent on the temperature. For instance, the exponentially increasing regime in the case of a chaotic billiard appears in a limited time-window that shrinks as the temperature increases \cite{Jalabert:2018da}. While most of the results concerning the initial growth of the OTOC were obtained through numerical calculations, there exist some analytic results. Among them, the short-time exponential growth-rate for the ``cat map" that has been analytically shown to be given by the classical Lyapunov exponent \cite{GarciaMata:2018kz}, and the semi-classical approaches that allowed to establish a connection with the classical Lyapunov exponent for a chaotic stadium billiard \cite{Jalabert:2018da} and for an interacting boson system \cite{rammensee2018many}.

As stated in the introduction, the long-time behavior of the OTOC in the chaotic case is characterized by a clear saturation  \cite{chen2018operator,rammensee2018many,GarciaMata:2018kz,Jalabert:2018da,hashimoto,cotler}. Unitarity, ergodicity and finite-size yield, for  definition \eqref{eq:OTOCinftemp} of the infinite temperature case, a limiting value of 1. In the general case of an initial thermal state, the limiting value depends on the chosen operators, and for the canonical choice of position and momentum operators, it is proportional to the temperature \cite{hashimoto,Jalabert:2018da}.

Classically integrable systems do not show a clear saturation of the OTOC, and the long-time limit is signed by strong oscillations. In the case of the square billiard, large periodic oscillations arising from the commensurability character of its energy spectrum prevent the approach to saturation in the long-time limit \cite{hashimoto}. The typical long-time behavior for the intermediate case of mixed systems is sketched in Fig.~\ref{Fig:OTOC_show}, where irregular oscillations are superposed to a saturation value. The main result of this paper is the characterization of these aperiodic oscillations in very different systems, and linking this information with the one stemming from other quantum and classical chaos indicators.

\section{\label{sec:chaos_in}Quantum Chaos Indicators}
We will characterize different mixed quantum systems with two widely used quantum chaos indicators in order to gauge the transition from integrability to chaos. The first indicator is the Brody parameter $\beta$, obtained by fitting the level-spacing distribution $P\left(s\right)$ with the Brody distribution \cite{Brody}, defined as
\begin{equation}
\small{P_{B}\left(s\right)=\left(\beta+1\right)b\,s^{\beta}e^{-b\,s^{\beta+1}}},\quad \small{b=\left[\Gamma\left(\frac{\beta+2}{\beta+1}\right)\right]^{\beta+1}},
\end{equation}
where $\Gamma \left(x\right)$ is the gamma function. The Brody distribution $P_{B}\left(s\right)$ approaches
to Poisson distribution $P_{P}\left(s\right)$ for $\beta\rightarrow 0$ and resembles the Wigner-Dyson (WD) distribution $P_{WD}\left(s\right)$ when $\beta\rightarrow 1$. 
Since the seminal works by Berry and Tabor \cite{Level_statistics2}  and  Bohigas, Gianoni, and Schmidt \cite{Bohigas}  it is by now well established that a Poisson distribution is associated to nonergodic, regular systems while a behavior resembling a Gaussian ensemble and characterized by a WD distribution is associated to quantum chaotic dynamics.

The second indicator that we consider is the localization of eigenstates, characterized by the inverse participation ratio (IPR). 
Suppose $\ket{\psi_i}$ is an eigenstate of the system of interest written in an arbitrary basis  $\{\ket{\phi_j}\}_{j=0}^{D-1}$ as 
$\ket{\psi_i}=\sum a_{ij} \ket{\phi_j}$.
We will denote the IPR of the eigenstate  as the inverse of the second moment of the distribution elements
\begin{equation}
\label{eq:def_ipr}
\xi_{E} \left(i\right) = \left(\sum^{D-1}_{j=0} \left|a_{ij}\right|^{4}\right)^{-1}.
\end{equation}
Therefore
$\xi_{E}$ is a measure of localization relative to the original basis and is defined as where small values of $\xi_{E}$ characterize a localized eigenstate while larger values signal delocalization. For systems with a WD distribution 
the coefficients $a_{ij}$ are independent random variables. These types of states are completely delocalized, having the direct consequence that $\xi_{E}^{\rm deloc} \approx D/3$, because the coefficients  $\left|a_{ij}\right|^{2}$ fluctuate \cite{LS_ipr,ZELEVINSKY199685}. In the numerical calculations we will consider the average over all the eigenstates 
\begin{equation}
\bar{\xi}_E = \frac{1}{D \ \xi_E^{\rm deloc}} \ \sum_{i=0}^{D-1}\xi_E(i).
\label{eq:xiE}
\end{equation}

{While the two previous indicators have been consistently employed in quantum chaos studies \cite{stockmann_1999}, they are known to exhibit some shortcomings. On the one hand, the measure \eqref{eq:xiE} is basis dependent. On the other hand, the Brody distribution is purely empirical, lacking a sound theoretical foundation. We therefore benchmark the previous indicators with two alternative measures, defined below.  

The Berry-Robnik (BR) distribution \cite{Berry_1984}, defined as 
\begin{equation}
\begin{split}
P_{BR}\left(s\right)= & \left[2\left(1 - \bar{\rho}\right)\bar{\rho}+\frac{\pi}{2}\bar{\rho}^{3}s\right]e^{-\left(1 - \bar{\rho}\right)s-\frac{\pi}{4}\bar{\rho}^{2}s^{2}} \\
& +\left(1 - \bar{\rho}\right)^{2}\text{erfc}\left(\frac{\sqrt{\pi}}{2}\bar{\rho}s\right)e^{-\left(1 - \bar{\rho}\right)s} \, ,
\label{eq:BRdistr}
\end{split}
\end{equation}
describes the gradual transition from Poisson-like to Wigner-like behavior under the evolution between integrable and completely chaotic regimes. While \eqref{eq:BRdistr} is not the most general BR distribution, it is accurate for the vast majority of systems (see Ref.~\cite{Berry_1984}). Here $\bar{\rho}^2$ plays the same role as the parameter $\beta$ does in the Brody distribution, and for the systems considered in our numerical work, both parameters are in qualitative agreement. It has been shown that, in the asymptotic semiclassical limit, the BR distribution represents an exact description for the spectral properties of the mixed dynamics \cite{Prosen_1994}, and that the fitted BR parameter coincides with the value obtained from the phase-space portions of the integrable and the nonintegrable parts \cite{stockmann_1999}. 

In the context of many-body systems, another quantity related to level spacing distributions can be used as a benchmark. The distribution of ${\rm min}(1/r,r)$ \cite{Oganesyan_2007,KazueTetsuo_2018,Atas_2013}, with $r$ being the ratio between the two nearest neighbor spacings of a given level, has the advantage of not requiring an energy unfolding thus, avoiding an important difficulty encountered in many-body systems, where the functional form of the level density is typically not known and there might not be enough statistics to extract it. The mean value $\overline{{\rm min}(1/r,r)}$ attains a minimum $I_P\approx 0.386$ when the statistics is Poissonian and a maximum $I_{WD}\approx 0.586$ when the statistics is Wigner-Dyson. Therefore, we can define the quantity

\begin{equation}
\eta\equiv\frac{\overline{{\rm min}(1/r,r)}-I_P}{I_{WD}-I_P} \, .
\end{equation}
Consistently with the $\beta$ and $\bar{\rho}^2$ parameters, $\eta\to 0 $ signals if the system is regular and $\eta\to 1$ if the system is chaotic.}
%
%
%
\section{Systems and results}
\label{sec:sysres}

\subsection{Overview}
\label{sec:Overview}

\begin{figure}
    \includegraphics[width=0.95\linewidth]{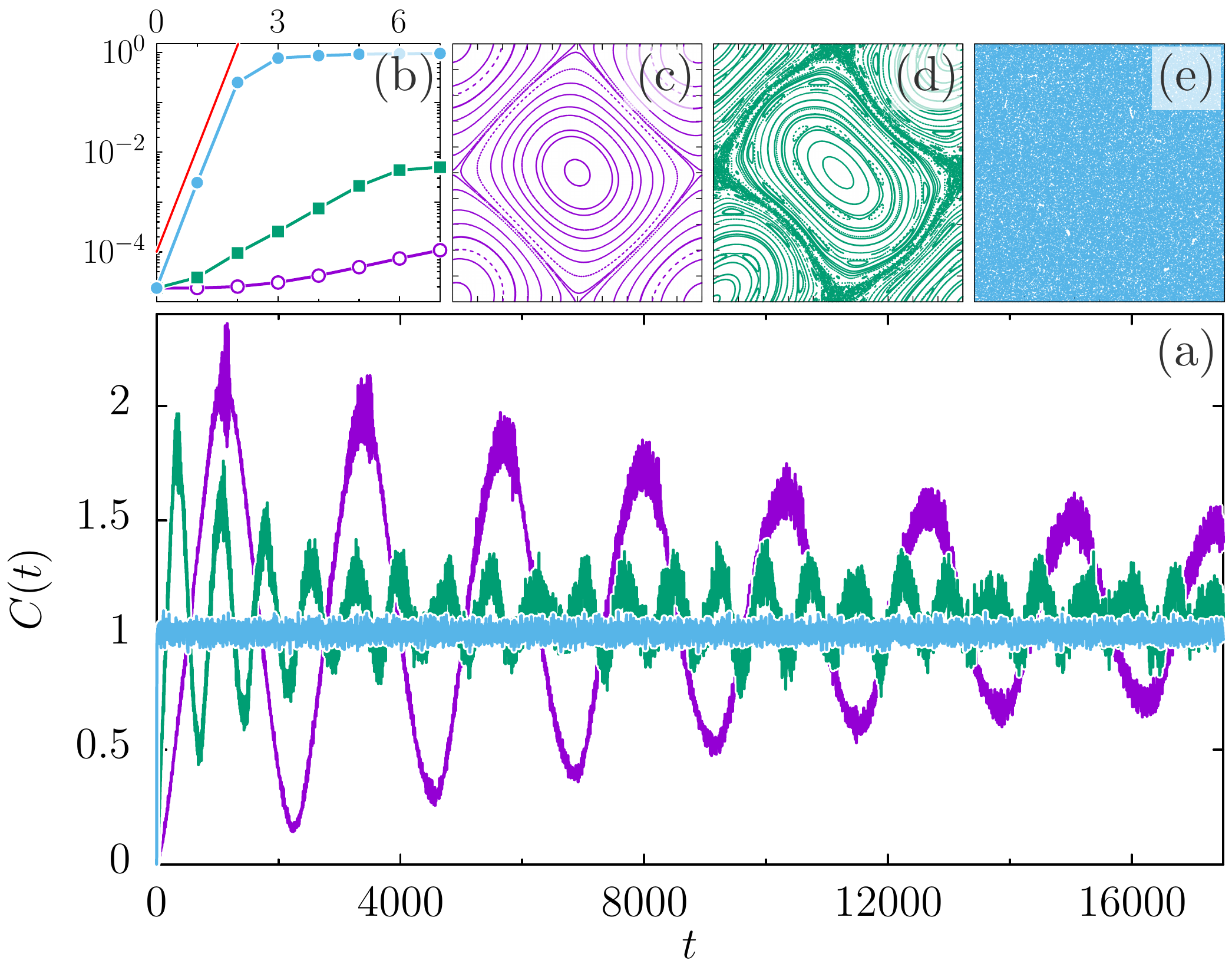}
    \caption{Main panel: (a) example of the typical behavior of $C(t)$ across a transition from integrability to chaos. The system is the Harper map described in SubSec.~\ref{sec:Harper}, having $K=0.063$ (large amplitude oscillations, violet), $0.19$ (medium amplitude oscillations, green), and $0.75$ (small amplitude oscillations, blue), with $D=200$. Insets: (b) Short-time of the OTOC for the previous values of $K$, but with $D=1024$. The same color code is used, and the symbols are open circles, squares, and circles, by increasing values of $K$. The red line shows $\sim e^{2\lambda_{\rm L} t }$; (c-e) Phase portraits for $K=0.063,\, 0.19,\,  0.75$ (from left to right) using the same color code as before.}
        \label{fig:sys_esquema}
\end{figure}

We start by presenting numerical studies of the OTOC in the long-time regime using a one-body system (the Harper map, whose precise description is given in Sec. \ref{sec:Harper}) in which the transition from integrability to chaos is clear from the available classical counterpart. In Fig.~\ref{fig:sys_esquema} such a transition is obtained by varying the parameter $K$ of the corresponding Hamiltonian, and represented by the color change from violet to green, and then to blue. Figures 2(c), 2(d), and 2(e) present the corresponding classical phase portraits as a Poincare surface of section for $K=0.063$, $0.19$, and $0.75$, respectively. Figures 2(a) and 2(b) show, respectively, the long- and short-time regimes of the OTOC $C(t)$ of the quantum map for the three different cases. For short times the growth of $C(t)$ is strongly dependent on $K$, and only in the completely chaotic case can an exponential growth with the Lyapunov exponent (red line) be identified, as it was shown in \cite{GarciaMata:2018kz}. For the very long times of the main panel $C(t)$ oscillates around the saturation value. In the integrable case a strong oscillatory behavior is observed (violet curve). The oscillations are characterized by a large amplitude and a seemingly small number of frequency components. The amplitude of the oscillations decreases as the chaos parameter $K$ becomes larger (green curve), approaching small quasirandom fluctuations around a constant saturation value in the fully chaotic regime (blue curve). The presented behavior is generic to other one-particle systems (data not shown). 

\begin{figure}
    \includegraphics[width=0.95\linewidth]{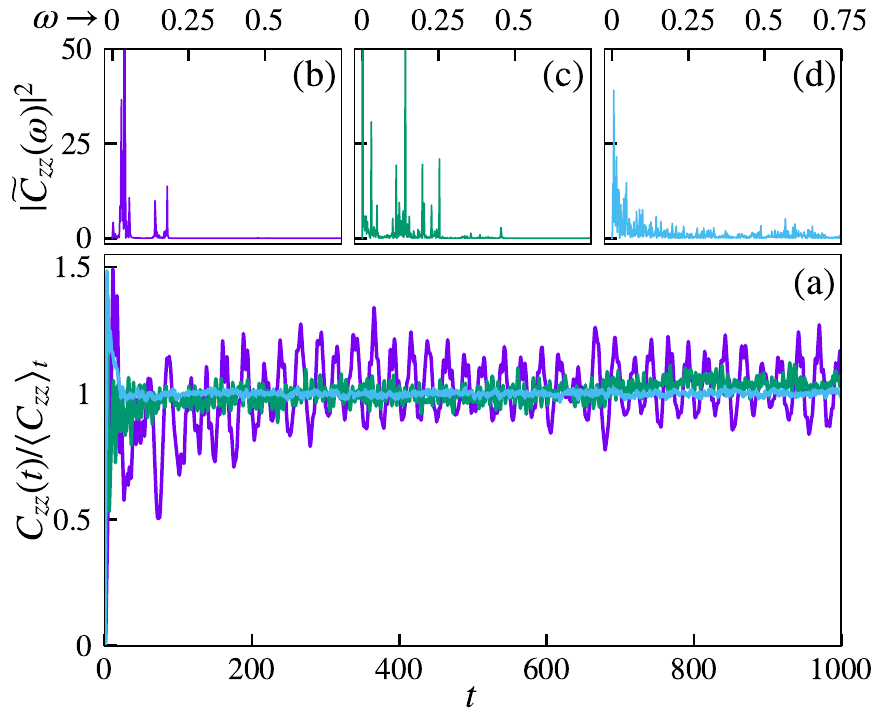} 
\caption{(color online). Main panel: (a) typical behavior of $C_{zz}\left(t\right)/\left\langle C_{zz}\right\rangle _{t}$ ($l=1$) across the transition from integrability to chaos for the Ising model with tilted magnetic field described in Sec.~\ref{subsec:tilted}. The chain length is $L=8$ ($D=256$). The chosen angles are $\theta=0.08\pi/2$ (large amplitude oscillations, violet), $0.31 \pi/2$ (medium amplitude oscillations, green), and $0.79\pi/2$ (small amplitude oscillations, blue). Upper panels: (b-d) normalized OTOC FFT distribution of frequencies $\left|\widetilde{C}_{zz}\left(\omega\right)\right|^{2}$, from the data of the main panel (a) for increasing values of $\theta$ (from left to right) using the same color code as before.}
    \label{fig:raw_otocs}
\end{figure}

The conclusions extracted from the previously discussed one-body example carry over to the many-body case. Figure~\ref{fig:raw_otocs}(a) shows the long-time behavior of the OTOC for an Ising chain with a tilted magnetic field (described in Sec.~\ref{subsec:tilted}). The transition from integrability to chaos is driven by the angle $\theta$ and represented by the color change from violet to green, and then to blue. Like in the one-body example, the oscillations around the saturation value decrease upon approaching the chaotic limit. The insets portray the magnitude square of the Fourier transform $\widetilde{C}\left(\omega\right)$, which helps to characterize the long-time OTOC oscillations and is used to define the $\xi_{_{OTOC}}$ parameter of Eq. \eqref{eq:xiOTOC}.  

\subsection{Long-time indicators for the OTOC}
\label{sec:LTIOTOC}

The generic behavior of the OTOC presented in the previous section lead us to conjecture that measuring and quantifying the oscillatory behavior of $C(t)$ in the long-time regime allow us to assess the chaotic nature of the quantum system. The suggested link between the integrability of a quantum system and the long-time oscillations of $C(t)$ makes it necessary to develop quantum indicators that are able to gauge the importance of these oscillations. 
 
A direct quantity to be used in order to characterize the oscillations is the standard deviation 
$\sigma_{_{OTOC}}=\sqrt{\langle C(t)^2\rangle-\langle C(t)\rangle^2}$, where the averages are taken over intermediate time-windows. A second useful quantity is the localization in Fourier space, obtained by computing  the inverse participation ratio of the Fourier transform $\widetilde{C}(\omega)$ of $C(t)$ through 
\begin{equation}
\xi_{_{OTOC}}=\left(\int_{0}^{\infty}d\omega|\widetilde{C}\left(\omega\right)|^{4}\right)^{-1} \, .
\label{eq:xiOTOC}
\end{equation}
To avoid the initial transient and a resulting large peak around $\omega=0$, we compute the Fourier transform of $C(t>t_0)-\left\langle C(t>t_{0})\right\rangle$. Examples of $\widetilde{C}\left(\omega\right)$ across the transition from integrability to chaos for the case of an Ising chain with a tilted magnetic field are given in the upper panels of Fig.~\ref{fig:raw_otocs}.

Just like the previously defined $\xi_{E}$, a small $\xi_{_{OTOC}}$ characterizes a very localized signal, meaning that a small number of frequencies are present, which is characteristic in the weakly chaotic chase. On the contrary a large $\xi_{_{OTOC}}$  corresponds to delocalization in frequency space and an almost constant value for $C(t)$.
In the following sections we test if the long-time regime of the OTOC can detect the quantum chaos transition in paradigmatic models of one and many-body systems.

We note that we will consider systems that depend on a parameter and we are
interested in the variation of $\xi_{_{OTOC}}$ and  $\sigma_{_{OTOC}}$ with the parameter. To compare both quantities behavior in a unique plot, we normalize them with respect to their maximum value present in the studied parameter ranges, $\bar{\sigma}_{_{OTOC}}^{-1}=\sigma_{_{OTOC}}^{-1}/(\sigma^{min}_{_{OTOC}})^{-1}$ and $\bar{\xi}_{_{OTOC}}=\xi_{_{OTOC}}/\xi^{max}_{_{OTOC}}$.
\subsection{One-body systems: Quantum maps}
\label{sec:qmaps}
Classical maps on the 2-torus are the simplest systems which can have all the essential features of chaotic motion.
Here we will consider quantum maps on the torus which are
the quantized counterparts of a classical canonical transformation corresponding to these classical maps \cite{berry1979quantum,Hannay1980}. 
The torus structure implies  periodicity in position and momentum. This periodicity results upon quantization in a
discrete Hilbert space of dimension $D$,  and an effective Planck constant 
$h_{\rm eff}=1/(2\pi D)$. 
Given  a classical map $M$ the corresponding quantum map $U_M$ is then a unitary operator with an  $D \times D$ matrix representation, and the time  evolution is given in discrete steps by  $U_M^t$, with $t$ an integer. 

Quantum maps have been extensively used to study quantum chaos \cite{DimaScholar} and  irreversibility \cite{DiegoScholar}. 
Moreover  there exist efficient quantum algorithms for many of the well known quantum maps 
\cite{Schack1998,Bertrand2001,Weinstein2002,Bertrand2004}, making them 
interesting test beds of quantum chaos in experiments using  quantum simulators.

The OTOC that we will consider for maps is
\begin{equation}
    C(t)=\frac{1}{D}{\rm Tr}\left([\hat{X}(t),\hat{P}]^2\right)
\end{equation}
where $\hat{X}(t)=(\hat{U}^\dagger)^t \,\hat{X}\,\hat{U}^t$, and 
\begin{equation}
    \hat{X}\equiv\frac{\hat{U}_S-\hat{U}_S^\dagger}{2i},\ \
        \hat{P}\equiv\frac{\hat{V}_S-\hat{V}_S^\dagger}{2i}
\end{equation}
are  Hermitian operators defined in terms of the unitary Schwinger shift operators \cite{schwinger}. If $\ket{q},\ \ket{p}$ are, respectively, position and momentum states related by $\langle p\ket{q}=e^{-2\pi i q p/D}$ (with $q,p=0,\ldots ,D-1$) , then
\begin{equation}
    \hat{V}_S\ket{q}=\ket{q+1},\ \hat{U}_S\ket{p}=e^{2\pi i/D}\ket{p}\,.
\end{equation}
In the semiclassical limit $\hat{X}$ and $\hat{P}$ approximate the position and momentum operators, respectively.

The two maps that we consider, described below, are derived from kicked systems, so they can be written as 
\begin{equation}
U=T(\hat p)V(\hat q).
\end{equation}
The advantage of the previous formulation is that the numerical implementation of the time evolution 
(and the corresponding diagonalization \cite{geisel}) 
becomes very efficient by using fast Fourier transformations.
For both of the maps that we consider, the kicking strength is the chaos parameter.
%
\subsubsection{Standard map}
The quantum (Chirikov) standard map (SM) \cite{DimaScholar} is defined by the evolution operator
\begin{equation}
\label{standardU}
U_{K}^{({\rm SM})}=e^{-i \frac{{\hat p}^2}{2\hbar}}e^{-i \frac{K}{\hbar} \cos(2 \pi \hat x) },
\end{equation}
and corresponds to the classical map
\begin{equation}
\begin{array}{lll}
p_{n+1}&=&p_n+\frac{K}{2\pi}\sin(2 \pi x_n)\\
x_{n+1}&=&x_n+p_{n+1}
\end{array}
\mod 1.
\end{equation}
For small values of $K$ the dynamics are regular. Below a certain critical value $K_c$ the motion in momentum is 
limited by KAM curves. These are invariant curves with irrational frequency ratio (or winding number) which represent quasiperiodic motion, and they  are the most robust orbits under nonlinear perturbations \cite{Licht}. At $K_c=0.971635\ldots$, the last KAM curve, with most irrational winding number, breaks down \cite{Greene1979}. Above $K_c$ there is unbounded diffusion  in $p$. For 
very large values of $K$, there might exist islands, but the motion is essentially chaotic.

\begin{figure}[h]
\centering
\includegraphics[width=0.95\linewidth]{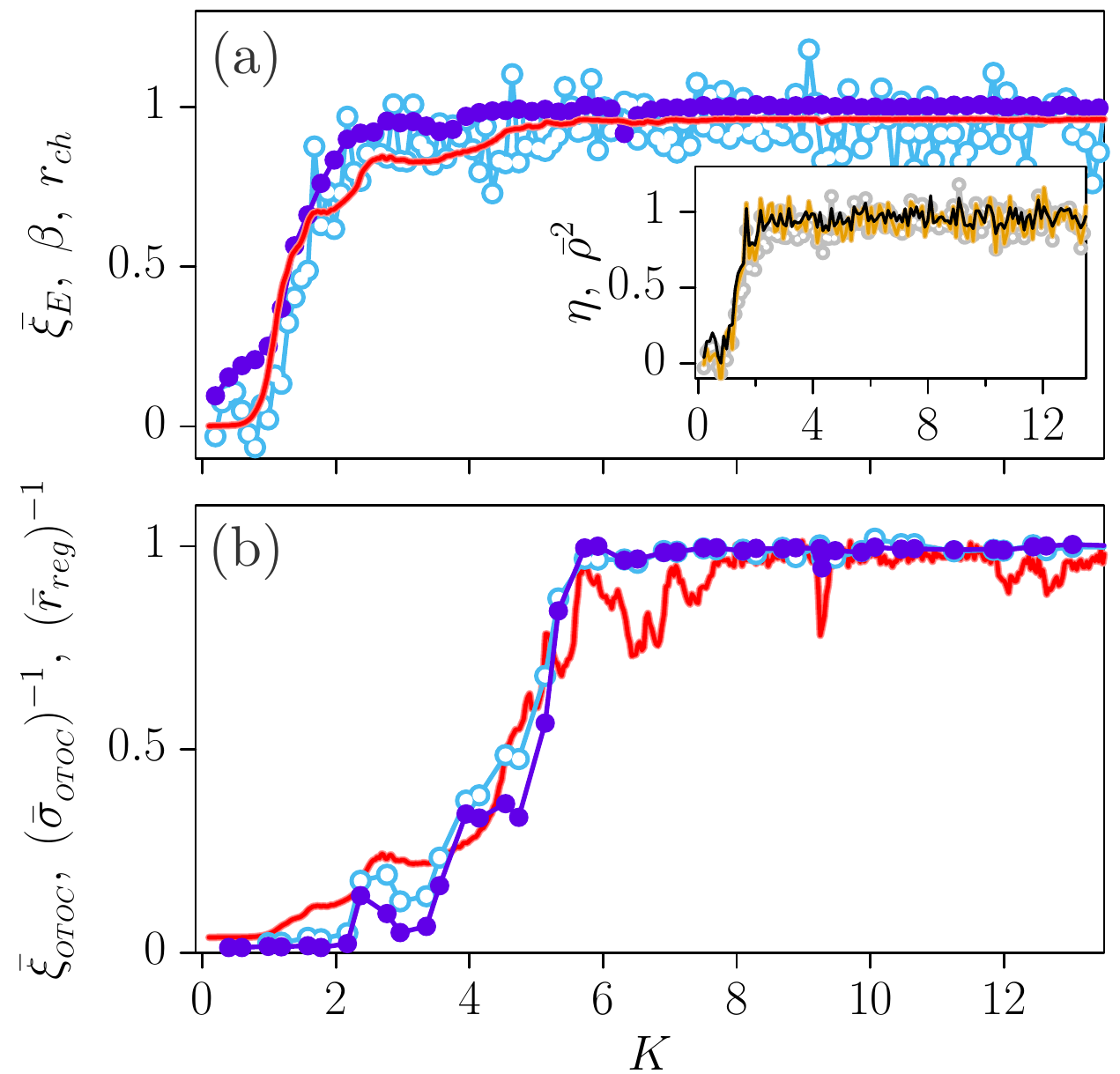}   
\caption{\label{fig:STD} (a) $\bar{\xi}_E$ (filled circles) and the $\beta$ parameter (open circles) resulting from the eigenvalues and eigenvectors of the Standard map with $D=1000$. The red solid line is $r_{ch}$ (top) obtained from the corresponding classical map, as described in App.~\ref{app:metro} with $n_{tot}=35000$ and averaging the curves obtained for $t_{max}=490,491,\ldots, 509$.
In the inset we show the values of $\bar{\rho}^2$ (black) obtained from the Berry-Robnik distribution and $\eta$ obtained from the ratios distribution. (b) Normalized $\bar{\xi}_{_{OTOC}}$ (filled circles) and $(\bar{\sigma}_{_{OTOC}})^{-1}$ (open circles) for $D=600$. The number of iterations is $6\times 10^3$. The red solid line corresponds to $1/r_{reg}= 1/(1-r_{ch})$, normalized as described in the text.
}
\end{figure}

In Fig.~\ref{fig:STD} we show the numerical results obtained for the standard map. In  {Fig.~\ref{fig:STD}(a)} the Brody parameter $\beta$ and $\bar{\xi}_{E}$  exhibit the same transition from localized (nonergodic) behavior  to delocalized (ergodic) behavior. The red curve is a Metropolis sampling approximation $r_{ch}$ of the area of the chaotic region for the classical map, described in Appendix~\ref{app:metro}. The direct correlation between these quantities is evident. {In the inset of the top panel of Fig.~\ref{fig:STD} {(a)} we present the corresponding curves for the parameter $\bar{\rho}^2$ obtained from the Berry-Robnik distribution (black). In gray we show the values of $\beta$ shown in the main panel. The qualitative behavior is evidently equivalent. Furthermore, we computed $\eta$  (orange) obtained from the ratios as defined in Sec.~\ref{sec:chaos_in}. It can also be seen that the behavior is completely equivalent to that of $\beta$.}

In {Fig.~\ref{fig:STD} (b)} we show the corresponding $\bar{\xi}_{_{OTOC}}$ and $(\bar{\sigma}_{_{OTOC}})^{-1}$. We use the relative variance [dividing $\sigma_{_{OTOC}}$ by the time average of $C(t)$] because this average changes with $K$ and because it reflects a relative deviation from the mean.
We can also see a transition from localized to delocalized  but not taking place at quite the same values arising from the quantum chaos indicators. The remarkable observation is that the OTOC indicators seem to reproduce the behavior of { $\bar{r}_{reg}^{-1}\equiv (r_{reg}/r_{reg}^{min})^{-1}$, the inverse normalized value of $r_{reg}=1-r_{ch}$ (defined in Appendix~\ref{app:metro}), which measures the area in the phase space of the integrable region ($r_{reg}^{min}$ corresponds to the saturation value of $r_{reg}$ at large $K$ )}. 
%
\subsubsection{Harper map}
\label{sec:Harper}
The quantum  Harper map (HM), defined by the evolution operator
\begin{equation}
U^{({\rm HM})}_{K_1,K_2}=e^{-i \frac{K_2}{\hbar} \cos(2 \pi \hat p)}e^{-i \frac{K_1}{\hbar} \cos(2 \pi \hat x) },
\label{harperU}
\end{equation}
is an approximation of the motion of kicked charge in the presence of an external magnetic field \cite{ArtusoScholar}. 
The corresponding classical map is 
\begin{equation}
\begin{array}{lll}
p_{n+1}&=&p_n-K_1\sin(2 \pi x_n)\\
x_{n+1}&=&x_n+K_2\sin(2 \pi p_{n+1})
\end{array}
\mod 1
\end{equation}
We will consider only the case  
$K_1 = K_2 = K$. For $K < 0.11$, the classical dynamics is regular, 
while for $K > 0.63$ it is fully chaotic for most values of $K$, although there are some particular values where small islands appear \cite{leboeuf1990}.

\begin{figure}[h]
\centering
\includegraphics[width=0.95\linewidth]{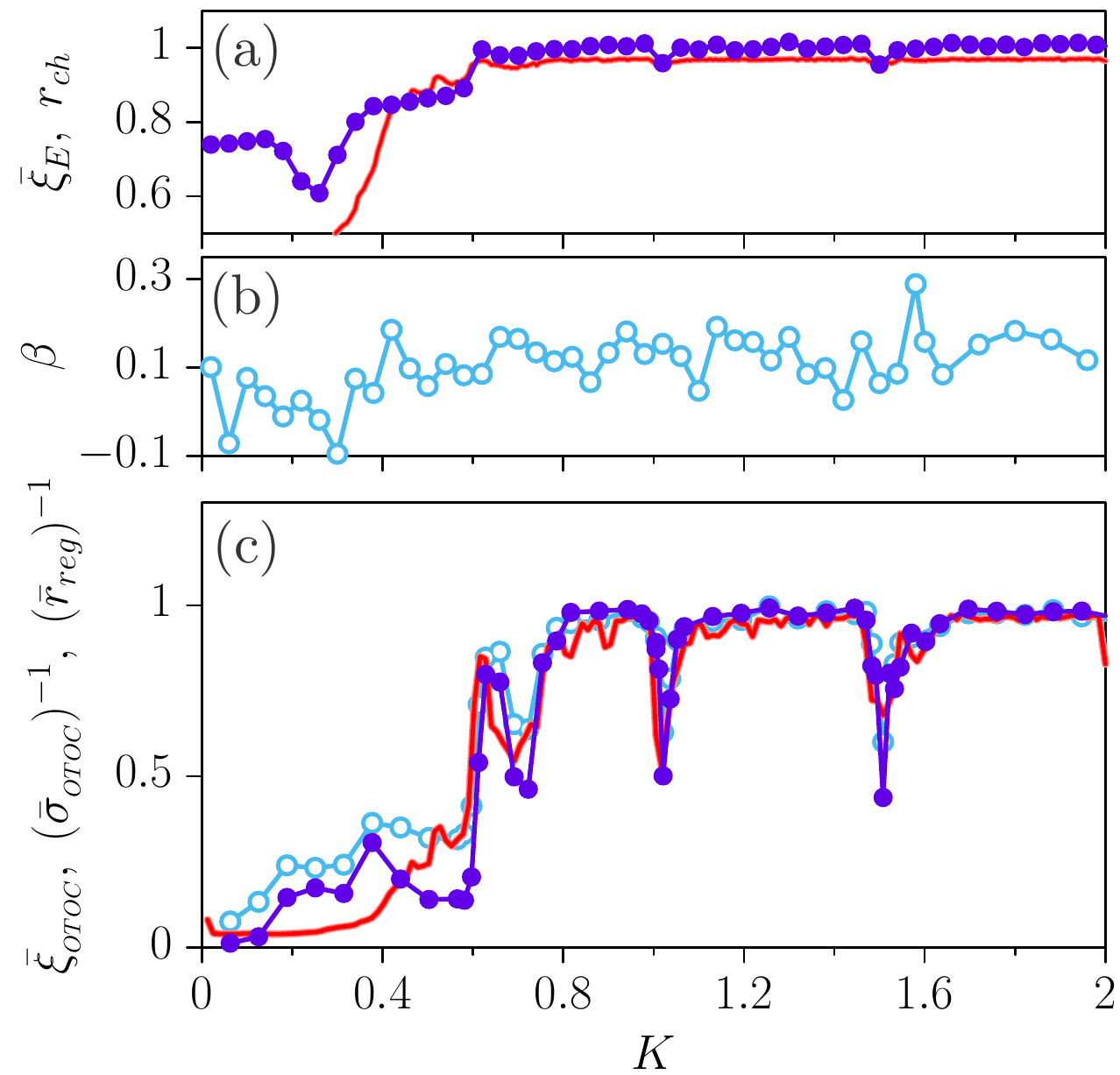}
\caption{\label{fig:HAR}
Same as Fig.~\ref{fig:STD} for the Harper map with $D=1000$ (a,b), and $D=200$ (c). The number of iterations is $2\times 10^4$. The red solid lines were obtained  using {$n_{tot}=35000$ and averaging the curves obtained for $t_{max}=90,91,\ldots, 109$.}}
\end{figure}

\begin{figure}[h]
\centering
\includegraphics[width=0.95\linewidth]{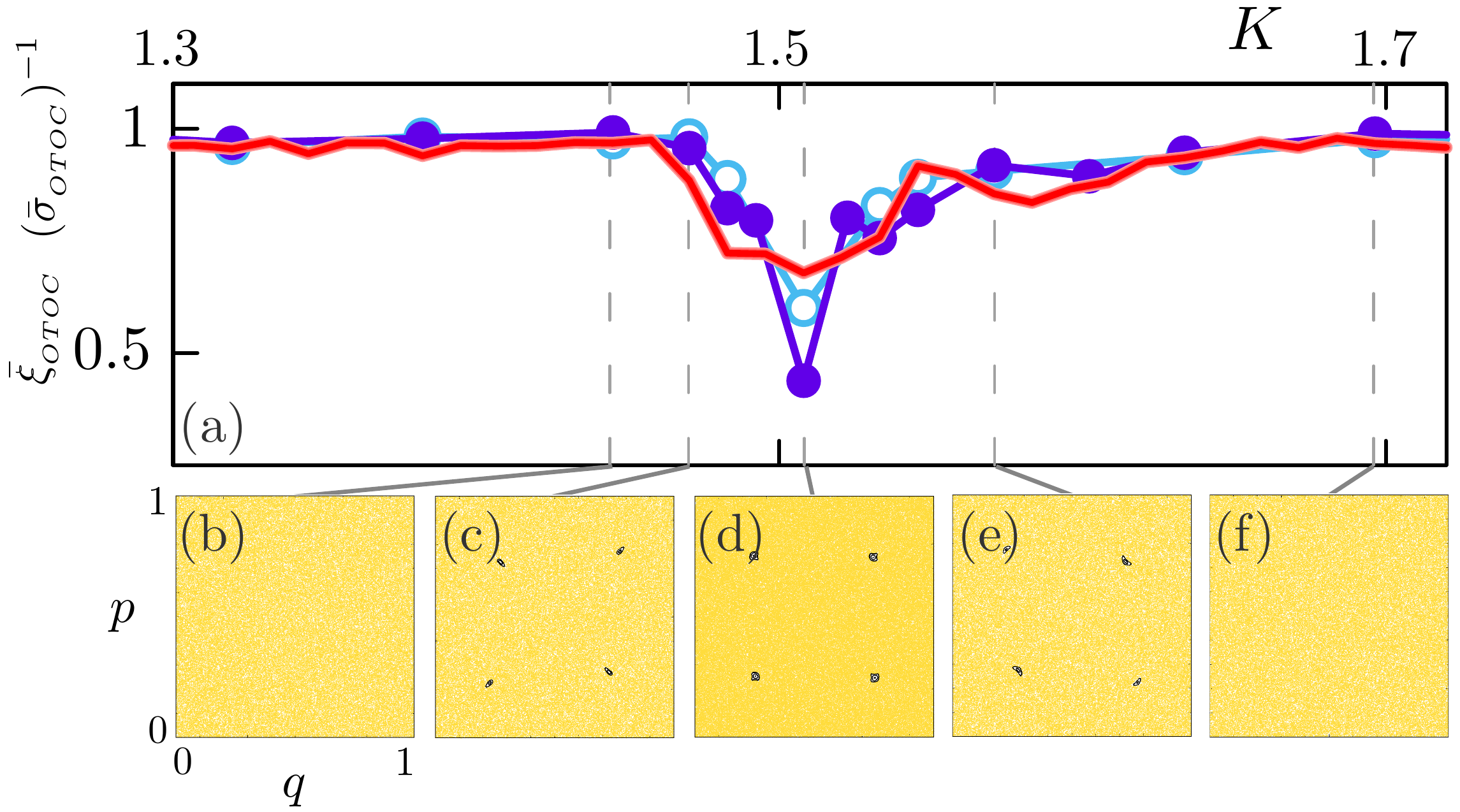}
\caption{\label{fig:picoHAR} {(a) Blowup of a section of  Fig.~\ref{fig:HAR} (c)} for values of $K$ around 1.5. Bottom panels: The phase portraits for the classical Harper map with { $K=1.44$ (b), $1.47$ (c), $1.508$ (d), $1.571$ (e) and $1.696$ (f)}. The lines indicate the position of the corresponding $K$ values on the axis in the top panel. Points inside regular islands are drawn darker to enhance contrast.}
\end{figure}

In Fig.~\ref{fig:HAR}{(b)} we see that for the Harper map $\beta$ does not change much with $K$. This is due to the fact that the map has symmetries, and it can be solved using an irrational $h=2\pi \hbar= 1/D$ (see Refs. \cite{borgonovi1995spectral,geisel}). For historical reasons we only consider a rational $h$, and therefore an approximately constant $\beta$ is obtained as expected. On the other hand a transition can be observed for the $\bar{\xi}_E$ {( Fig.~\ref{fig:HAR} {(a)})}.  As shown for the standard map, the classical chaotic area $r_{ch}$ closely follows the behavior of $\xi_E$ (red curve).

In {Fig.~\ref{fig:HAR} {(c)}} we see that the same qualitative behavior can be observed for $\bar{\xi}_{_{OTOC}}$ and $(\bar{\sigma}_{_{OTOC}})^{-1}$, obtained from the OTOC data  like the one presented in Fig.~\ref{fig:sys_esquema}. Similar to what happens with the standard map, a transition is visible, taking place at a value somewhat different from the one suggested by $\xi_E$. However, as noted above, $\bar{\xi}_{_{OTOC}}$ and $\bar{\sigma}_{_{OTOC}}$ follow surprisingly well the behavior of $\bar{r}_{reg}$, which implies that we can relate their behavior to the size of the regular islands in the corresponding classical phase space. 

It is known that for the classical Harper map, upon increasing $K$, there are values of for which there appear regular islands and then disappear. Such an effect translates into dips of $r_{reg}$. It is interesting to see, as is shown in detail in Fig.~\ref{fig:picoHAR}, that the $K$-dependences of $\bar{\xi}_{_{OTOC}}$ and $\bar{\sigma}_{_{OTOC}}$ reproduce the shape of the dip accurately. Thus, the proposed OTOC indicators seem to be well-suited to  identify  chaotic regions and describe a mixed phase space regime. 
\subsection{\label{sec:SCN}Many-body systems: Spin chains}
We consider three many-body spin-1/2 systems described by a Hamiltonian that depends on a tunable parameter (e.g. interaction strength). By changing this parameter the system is driven through a transition between integrable and chaotic regimes. From a quantum chaos point of view, the  transition can be observed through the spectra of eigenvalues or in the eigenstate distributions.
The chosen systems have been extensively used in studies of quantum thermodynamics \cite{PhysRevE.87.012118,PhysRevLett.106.040401,PhysRevLett.111.197203,PhysRevB.91.155123} and many-body localization \cite{Imbrie2016,PhysRevLett.109.017202,PhysRevLett.117.027201}. 

In the following discussion $\hbar$ is set to $1$, while $L$ refers to the number of spin-1/2 sites in the chain and 
\begin{equation}
\hat{S}_{i}^{\mu}=\frac{1}{2}\hat{\sigma}_{i}^{\mu} 
\end{equation}
are the spin operators at site $i=0,1,...,L-1$, with $\hat{\sigma}^{\mu}$ the corresponding Pauli matrix associated with the direction $\mu=x,y,z$. Boundary conditions for all the spin chain models are set as open. Since the spin operators are both unitary and Hermitian, 
the OTOC can be written for infinity temperature as
\begin{align}
C_{\mu\nu}\left(l,t\right)&=\frac{1}{2}\left\langle [\hat{\sigma}_{0}^{\mu}(t),\hat{\sigma}_{l}^{\nu}\right]^{2}\rangle\nonumber\\
&=1-Re \left\{ Tr[ \hat{\sigma}_{0}^{\mu}(t)\hat{\sigma}_{l}^{\nu}\hat{\sigma}_{0}^{\mu}(t)\hat{\sigma}_{l}^{\nu}] \right\}/D,
\label{eq:OTOC_spin}
\end{align}
where $D$ is the dimension  of the  Hilbert space.
\subsubsection{Perturbed XXZ model}
 The first model we consider is an anisotropic spin-1/2 chain with nearest-neighbor (NN) interactions and a perturbation consisting in next-nearest-neighbor (NNN) interactions (tuned by a strength parameter $\lambda$). 
The Hamiltonian of the chain is
\begin{equation}
\hat{H}(\lambda)=\hat{H}_{0}+ \lambda \hat{H}_{1}
\end{equation}
with  
\begin{equation}
\hat{H}_{0}=\sum_{i=0}^{L-2}(\hat{S}_{i}^{x}\hat{S}_{i+1}^{x}+\hat{S}_{i}^{y}\hat{S}_{i+1}^{y}+\mu \hat{S}_{i}^{z}\hat{S}_{i+1}^{z}),
\label{Hcero}
\end{equation}
\begin{equation}
\hat{H}_{1}=\sum_{i=0}^{L-3}(\hat{S}_{i}^{x}\hat{S}_{i+2}^{x}+\hat{S}_{i}^{y}\hat{S}_{i+2}^{y}+\mu \hat{S}_{i}^{z}\hat{S}_{i+2}^{z}).
\end{equation}
This system has been extensively studied from the quantum chaos point of view in Ref. \cite{Borgonovi}. It presents a chaotic regime when the NNN coupling strength $\lambda$ becomes comparable with the NN coupling, turning the level-spacing distribution from Poisson to WD. The latter transition occurs when all symmetries are removed.
For this reason, in our calculations,  the parameter $\mu$ is fixed at $0.5$, in order to avoid the conservation of total spin $\hat{S}^{2}$, which occurs at $\mu=1$.
The $z$-component of spin $\hat{S}^{z}=\sum_{i=0}^{L-1}\hat{S}_{i}^{z}$ is conserved in our model.
This symmetry allows separation of the total spanned space, of dimension $D$, into smaller subspaces $\mathcal{\hat{S}}_{N}$ with a fixed number $N$ of spins up or down. 
The dimension of each subspace $\mathcal{\hat{S}}_{N}$ is 
\begin{equation}
    D_{N} \equiv {\rm dim} \left( \mathcal{\hat{S}}_{N} \right) = \begin{pmatrix}L\\
N
\end{pmatrix} = \frac{L!}{N!\left(L-N\right)!}.
\end{equation}

The system also presents a symmetry with respect to the parity operator $\hat{\Pi}$, defined through the permutation operators $\hat{P}_{ij}$ as
\begin{equation}
\hat{\Pi} = \hat{P}_{0,L-1}\hat{P}_{1,L-2}...\hat{P}_{\frac{L-1}{2}-1,\frac{L-1}{2}+1}.
\end{equation}
The defined operator $\hat{\Pi}$ is described for a spin chain of odd length $L$ (the even $L$ situation is analogous). The conservation of $\hat{\Pi}$ divides the spanned space into even and odd subspaces with dimensions $D=D^{Even}+D^{Odd}$, where $D^{Even/Odd}\approx D/2$. 

Similarly to the case presented in the previous section of quantum maps, we now study how the effect of the integrability to chaos transition, occurring in the eigenstates and eigenvalues spectral fluctuations, translates into the long-time properties of the OTOC. 

\begin{figure}[h]
\centering
\includegraphics[width=0.99\linewidth]{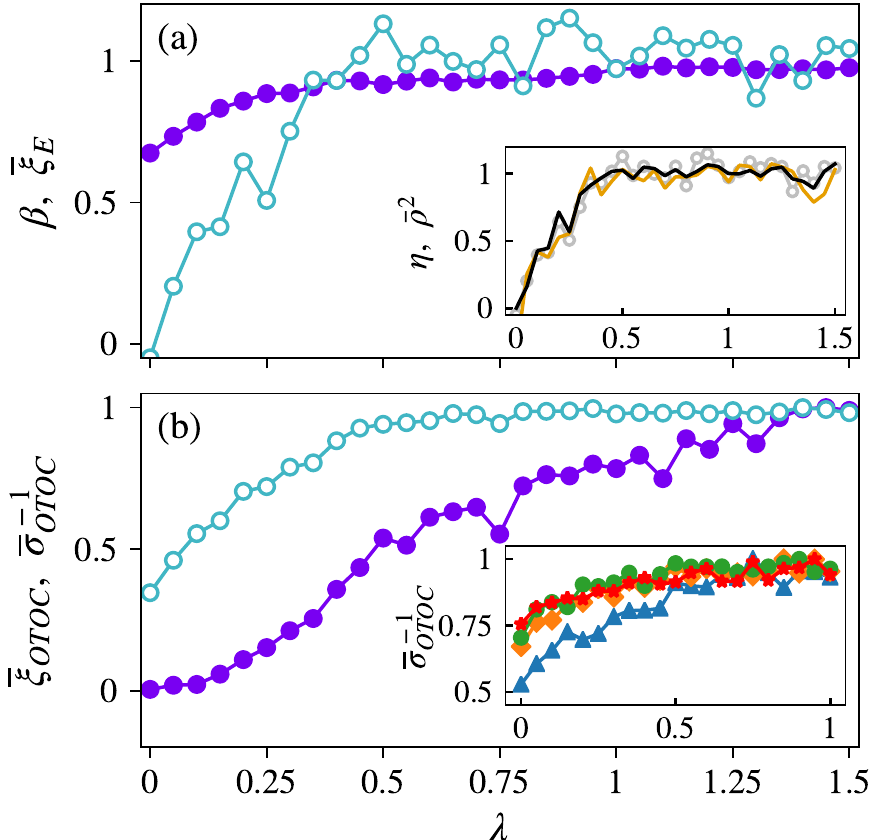} 
\caption{Chaos transition for the perturbed $XXZ$ spin chain. {(a)}: $\bar{\xi}_{_{E}}$ (filled circles), Brody parameter $\beta$ (open circles) for the even parity subspace in a spin chain of length $L=13$ and $N=5 $ $(D^{Even}=651)$. {{Inset (a)}: The values of $\bar{\rho}^2$ (black) and $\eta$ (orange) for the previous system conditions.} {(b)}: $\bar{\xi}_{_{OTOC}}$ (filled circles) and $\bar{\sigma}_{_{OTOC}}^{-1}$ (open circles) for a spin chain of length $L=13$ and $N=5$ (the entire Hilbert space has $D=1287$) and $l=1$. {Inset (b)}: $\bar{\sigma}_{_{OTOC}}^{-1}$ for a spin chain of length $L=10$ and $N=6$ (the full Hilbert space has $D=210$) for $l=1\,\text{(triangles)},\,2\,\text{(diamonds)},\,3\,\text{(circles)}$ and $4\,\text{(asterisks)}$.}
\label{Fig:BNVI_FFT}
\end{figure}

In {Fig.~\ref{Fig:BNVI_FFT}(a)} we show 
such a transition through the quantum chaos indicators 
$\beta$ and 
$\bar{ \xi}_{E} $ as a function
of the NNN coupling strength $\lambda$. In this calculations we analyzed a chain of length $L=13$ and $N=5$, where the even parity subspace has $D^{Even}=651$ states. Although it is more pronounced for $\beta$ than for $\bar{\xi}_E$, we clearly see in both measures that a transition occurs between
$\lambda = 0.3$ and $\lambda = 0.5$. The inverse participation ratio 
is computed with respect to the spin site basis and averaged over $10\%$ of the values in the center of the energy spectrum. {The inset of Fig.~\ref{Fig:BNVI_FFT}  (a) shows that both parameters $\eta$ and $\bar{\rho}^2$ exhibit similar behavior with respect to $\beta$.}

In {Fig.~\ref{Fig:BNVI_FFT}(b)} we present the measures 
$\bar{\sigma}_{_{OTOC}}$ and $\bar{\xi}_{_{OTOC}}$ characterizing the long-time behavior of the OTOC
$C_{zz}\left(l,t\right)$. 
In this case, the separation of the spin operators sites is $l=1$, but similar results were obtained for other separation values $l$.
In the {inset of Fig.~\ref{Fig:BNVI_FFT}  (b)} we show $\bar{\sigma}_{_{OTOC}}^{-1}$ as a function of the parameter  $\lambda$ for $l=1,2,3$ and $4$, for a small spin chain ($L=10$ and $N=6$).  We observe that the same qualitative behavior as in $\beta$ and $\bar{\xi}_E$, i.e. a transition as a function of $\lambda$, is observed for $\bar{\sigma}_{_{OTOC}}^{-1}$, showing that it is a good indicator of an integrable to chaotic transition. {For $\bar{\xi}_{_{OTOC}}$ the transition can be seen in the change of curvature that occurs at the same point that $\bar{\sigma}_{_{OTOC}}^{-1}$ reaches the saturation value.}  
We remark that in the case of the OTOC such a transition is already revealed for much smaller chain lengths than the ones needed to observe it for $\beta$ and $\bar{\xi}_E$.

\begin{figure}[h]
\centering
\includegraphics[width=0.99\linewidth]{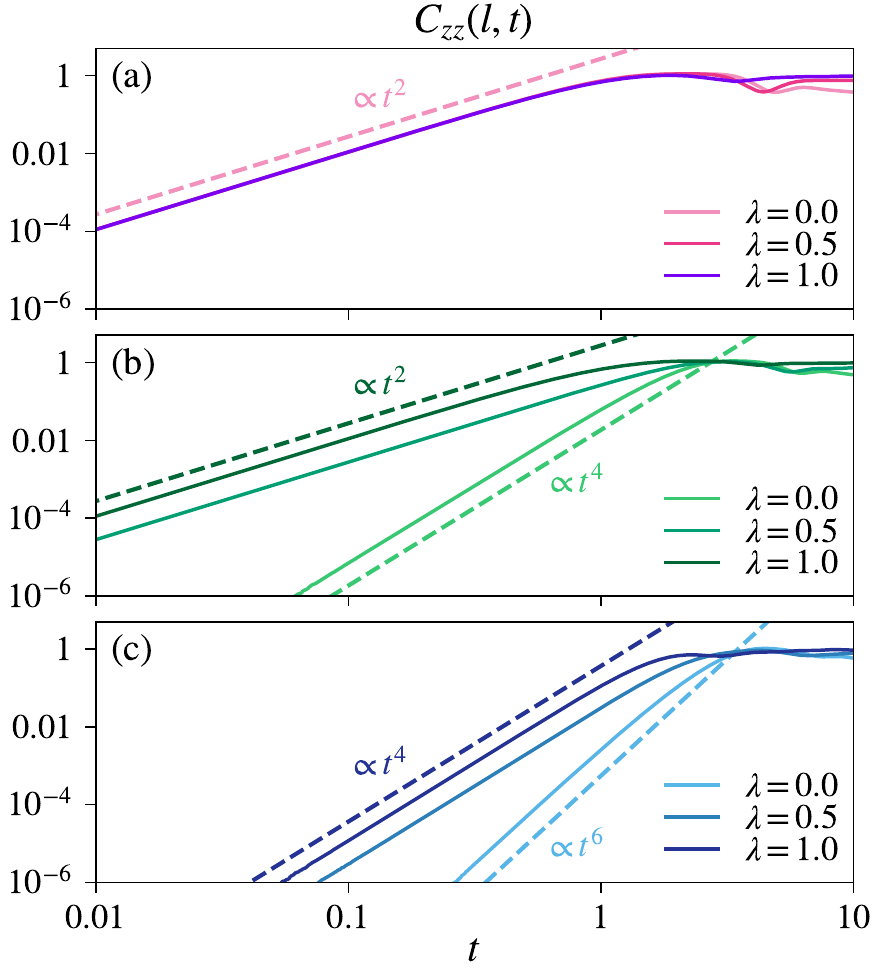} 
\caption{Short-time growth of $C_{zz}\left(l,t\right)$ for the perturbed $XXZ$ model with $L=9$, $N=5$, different spin separations { $l=1$ (a), $2$ (b), $3$ (c)} (solid lines) and different values of $\lambda=0,\,0.5,\,1$. Short-time power-law growth predicted with the HBC formula of Eq. (\ref{shorttime1}) is also shown (dashed lines).}
\label{Fig:BNVI_ST}
\end{figure}
We end the analysis of the perturbed $XXZ$ spin chain considering the short-time growth of the OTOC.
In Fig.~\ref{Fig:BNVI_ST} we show such a regime for different spin separations { $l=1$ [Fig.~8(a)], $2$ [Fig.~8(b)], $3$ [Fig.~8(c)]} and 
perturbation strength $\lambda$. We can clearly see that the behavior is characterized by the power law (\ref{shorttime1}) that is obtained in Appendix~A using the HBC formula. 
As is evident from Eq.~(\ref{shorttime1}) and the data shown in   Fig.~\ref{Fig:BNVI_ST},
the short-time power-law growth is strongly dependent on the coupling strength $\lambda$, except for $l=1$, where the initial growth is independent of $\lambda$. 
We remark that this short-time regime is not influenced by the integrable to chaotic transitions
shown in the top panel of Fig.~\ref{Fig:BNVI_FFT}.

\subsubsection{Ising model with tilted magnetic field}\label{subsec:tilted}
The second model consists of a spin-1/2 chain with NN interactions (Ising model) with a tilted magnetic field. The Hamiltonian of the system is
\begin{equation}
\hat{H}(\theta)=J\sum_{i=0}^{L-2}\hat{S}_{i}^{z}\hat{S}_{i+1}^{z}+\small{B}\sum_{i=0}^{L-1}\left(\sin\left(\theta\right)\hat{S}_{i}^{x}+\cos\left(\theta\right)\hat{S}_{i}^{z}\right).
\end{equation}
Parameters are set at $J=2$ and $B = 2$. When the angle $\theta = 0$, the magnetic field is longitudinal and when $\theta=\pi /2$, it becomes the transverse field model. In both cases the system is integrable with a highly degenerate spectrum. The Jordan-Wigner (JW) transformation \cite{sachdev2001quantum} yields the solution of a noninteracting fermionic system. At intermediate angles $0<\theta<\pi/2$, the model undergoes a quantum chaos transition. 
In this case, JW transformation maps the system to a model of interacting fermions. 
The quantum chaos transition and eigenvalues spectral properties have been studied in Ref.~\cite{tilt}, in our analysis a WD NN level-spacing distribution occurs near $\theta=\pi/4$.
\begin{figure}[h]
\centering
\includegraphics[width=0.99\linewidth]{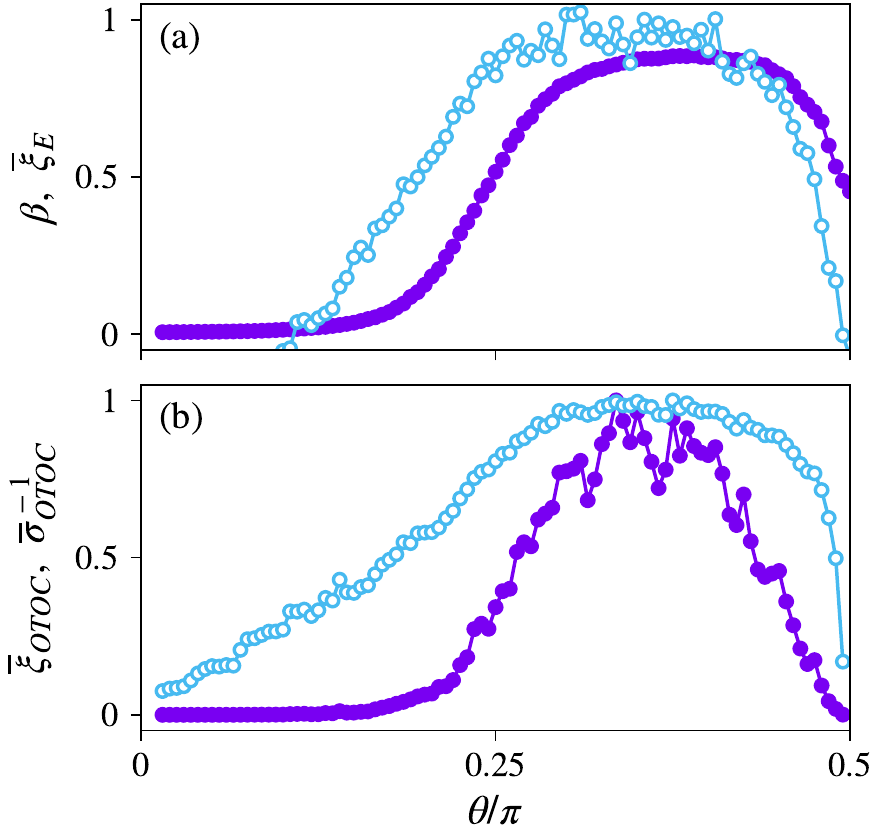} 
\caption{Chaos transition for the Ising model with tilted magnetic field. {(a)}: $\bar{\xi}_{_{E}}$ (filled circles) and Brody parameter $\beta$ (open circles) for the even parity subspace for a spin chain length $L=12$ $(D^{Even}=2080)$. {(b)}: $\bar{\xi}_{_{OTOC}}$ (filled circles) and $\bar{\sigma}_{_{OTOC}}^{-1}$ (open circles) for a spin chain of length $L=8$ $(D=256)$. It is important to remark that the plot begins at $\theta=0.03\pi/2$.}
\label{fig:Tilt_FFT}
\end{figure}

The parity symmetry described in the previous model is also present in this system and therefore, even and odd subspaces should be analyzed separately for the eigenstate and eigenvalue spectral properties. On the contrary, the OTOC analysis will be carried on the entire Hilbert space of dimension $D=2^{L}$. It is important to notice that at $\theta=0$ we have $C_{zz}\left(l,t\right)=0$, and as $\theta$ increases, the long-time values of the OTOC keep increasing. Therefore, since our interest stands in the OTOC oscillations with respect to its mean value, both $\bar{\xi}_{_{OTOC}}$ and $\bar{\sigma}_{_{OTOC}}^{-1}$ are analyzed for the scaled correlator $C_{zz}\left(l,t\right)/\left\langle C_{zz}\left(l\right)\right\rangle _{t}$, where $\left\langle C_{zz}\left(l\right)\right\rangle _{t}$ is the temporal average after the OTOC reaches its mean saturation value (see Fig.~\ref{fig:raw_otocs}). {It is important to remark that the smaller oscillations of the OTOC present in this spin chain (the ones we are interested in) were mounted over a smooth function over time for certain values of $\theta$, and therefore {those functions} were removed to obtain clear results for the long time indicators. }

\begin{figure}[h]
\centering
\includegraphics[width=0.99\linewidth]{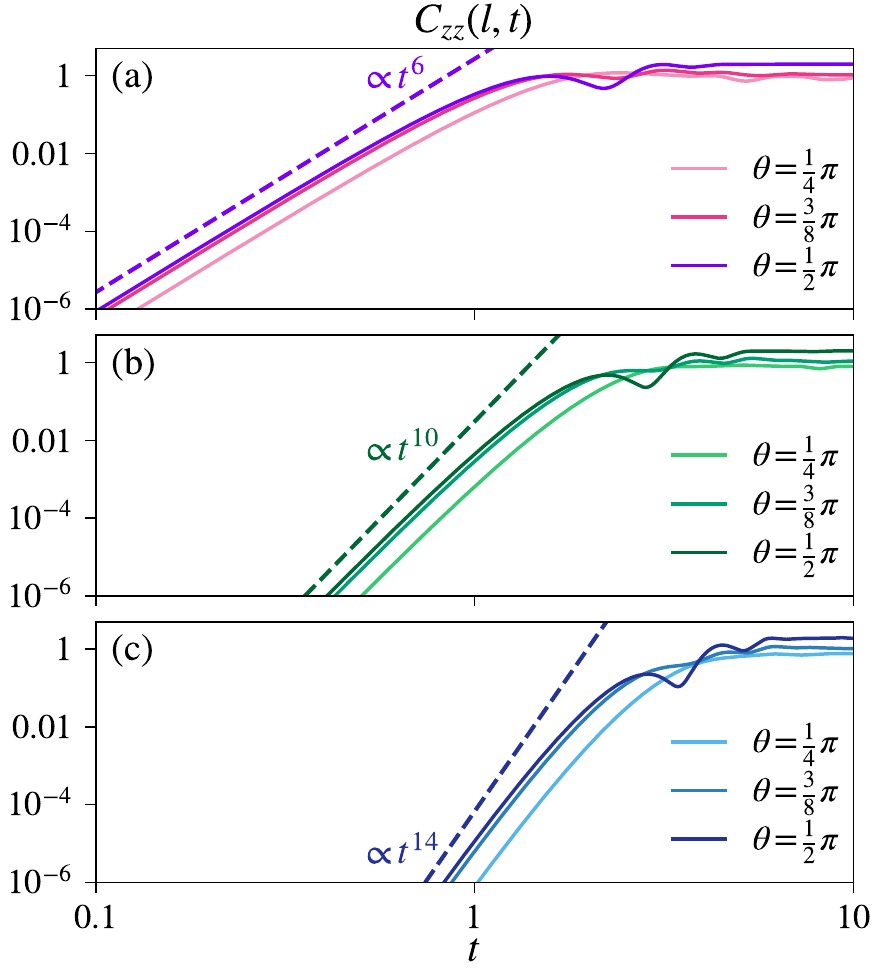} 
\caption{Short-time growth of {scaled} $C_{zz}\left(l,t\right)$ for an Ising chain of length $L=8$ subjected to a tilted magnetic field with different angles $\theta$ and various spin separations sites { $l=1$ (a), $2$ (b), $3$ (c)} (solid lines). The short-time power-law growth predicted by Eq. (\ref{stilt}) is also plotted (dashed lines).}
\label{fig:Tilt_ST}
\end{figure}
In {Fig.~\ref{fig:Tilt_FFT} (a)} we analyze the quantum chaos transition in the spectral properties and in
the long-time regime of the OTOC. In the top panel we show
$\beta$ and  $\bar{\xi}_{E}$
as a function of the tilt angle $\theta$ for a spin chain of length $L=12$. The calculations were 
done with the even subspace which has $D^{Even}=2080$ states. As we have previously noted, the system is integrable
for $\theta=0$ and $\pi/2$, where $\beta(0)$ takes negative values since the NN distribution resembles more a $\delta$ distribution than a Poisson one and $\beta(\pi/2)\approx0$. The WD distribution is reached at
$\theta\approx\pi/4$ where $\beta=1$.

In {Fig.~\ref{fig:Tilt_FFT} (b)} we show $\bar{\xi}_{_{OTOC}}$ and $\bar{\sigma}_{_{OTOC}}^{-1}$ for a $L=8$ spin chain length. The choice such that the lower panel uses a very small chain is intentional to highlight an interesting property. 
Not only do both results show nearly identical behavior concerning the transition to chaos, but the choice of a very small spanned Hilbert space dimension also highlights the fact that the OTOC does not require Hilbert spaces as large as those required by the statistical studies of eigenstate and eigenvalue properties. 
We checked that qualitatively equivalent results for larger spin chains can be obtained (data not shown).

In {Fig.~\ref{fig:Tilt_ST}} we show the short-time behavior of the OTOC $C_{zz}\left(l,t\right)$ for separation sites { $l=1$ [Fig.~10(a)], $2$ [Fig.~10(b)], $3$ [Fig.~10(c)]} and angles $\theta=1/4 \pi,\,3/8 \pi$ and $\pi/2$ of the magnetic field. The angle $\theta=\pi/2$ has been thoroughly studied in Ref.~\cite{Rusos}, and the short-time power-law formula presented in that work gets barely modified by the presence of angle $\theta$. The relation obtained by the HBC formula Eq.~(\ref{stilt}) is also plotted and clearly describes the short-time regime. 
As shown in Fig.~\ref{fig:Tilt_ST}, although angle dependence is present in Eq. (\ref{stilt}), it does not affect the short-time power law.
Furthermore, like in the previous model, the transition to chaos does not affect the short-time growth in any noticeable way.

%
%
\subsubsection{Heisenberg spin chain with random magnetic field}
The third model that we analyze consists of a spin chain with NN interactions [the Hamiltonian of
Eq.~(\ref{Hcero}) with $\mu=1$] coupled  with an external random magnetic field in the $z$-direction \cite{Marko2008}. The  Hamiltonian of the system is
\begin{align}
\hat{H} &= \sum_{i=0}^{L-2}( \hat{S}_i^x \hat{S}_{i+1}^x + \hat{S}_i^y \hat{S}_{i+1}^y  + \hat{S}_i^z \hat{S}_{i+1}^z)+\sum_{i=0}^{L-1} h_i \hat{S}_i^z.
\end{align}
where $h_i$ are independent random variables at each site, uniformly distributed in the interval $[-h,h]$.
This is a paradigmatic prototype model that has been used to study the many-body localization (MBL) transition. \cite{Pal2010,de2013ergodicity,bauer2013area,Luitz}
The transition in the level-spacing statistics has been a subject of study for quite some time  \cite{Avishai2002,santos2004integrability,Santos2005,Luitz}.
For $h=0$ and taking into account symmetries, it can be shown that the system is solvable and the nearest level-spacing distribution is Poissonian. As $h$ increases the disorder breaks the symmetries and the system starts to become chaotic, reaching a Wigner-Dyson distribution at 
$h\approx 0.5$, corresponding to ergodicity. Finally if the disorder is too strong there is a MBL transition (for a review see Ref.~\cite{ALET2018}).
\begin{figure} 
\centering
\includegraphics[width=0.99\linewidth]{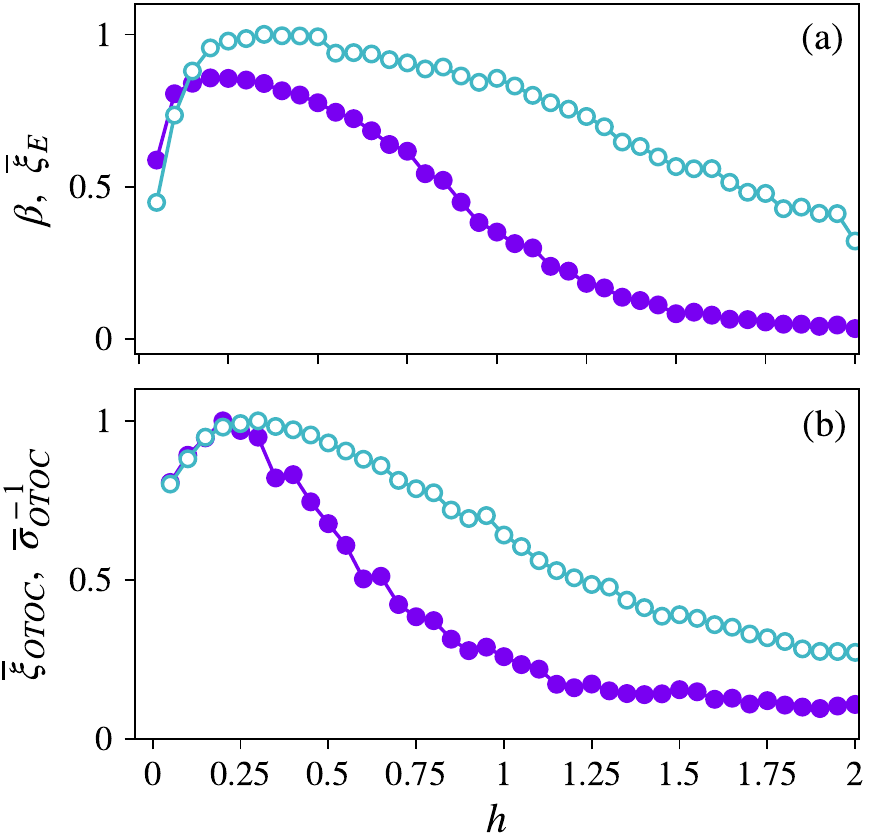}  
\caption{Chaos transition for the Heisenberg spin chain with random magnetic field. {(a)}:  $\bar{\xi}_{_{E}}$ (filled circles) and Brody parameter $\beta$ (open circles) in a chain of length $L=13$, $N=5$ $(D=1287)$ and averaged over 100 realizations. {(b)}: $\bar{\xi}_{_{OTOC}}$ (filled circles) and $\bar{\sigma}_{_{OTOC}}^{-1}$ (open circles) for a chain of length $L=9$ and $N=5$ $(D=126)$ and averaged over 100 realizations. The plot begins at $h=0.05$.}
\label{Fig:LS_FFT}
\end{figure}

\begin{figure}[h]
\centering
\includegraphics[width=0.99\linewidth]{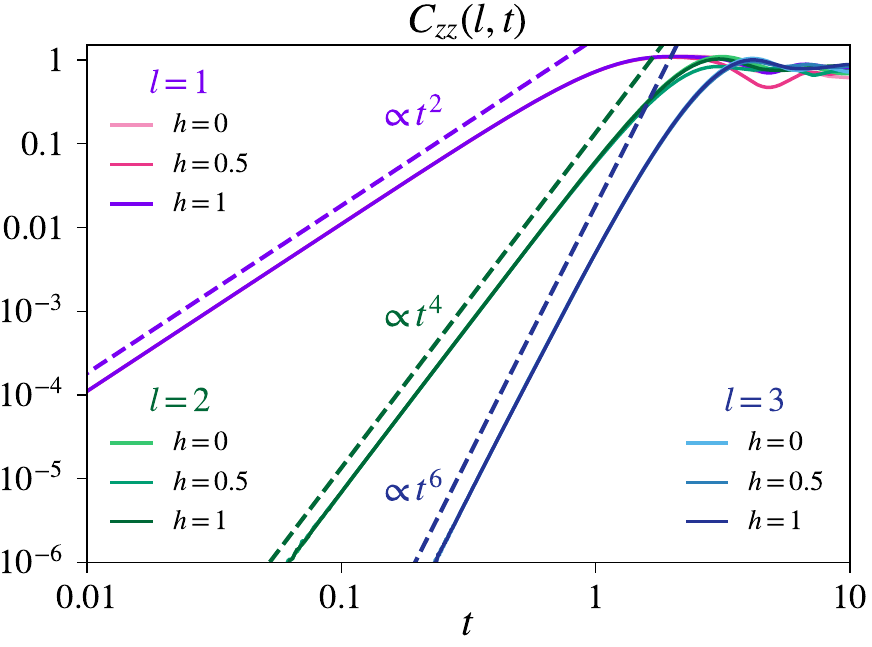} 
\caption{Short-time growth of $C_{zz}\left(l,t\right)$ for the Heisenberg spin chain with random magnetic field and a chain of length $L=9$, $N=5$ for different spin separations $l=1,2,3$ (solid lines). Short-time power-law growth is predicted with the HBC formula \ref{short-hei} (dashed lines).}
\label{Fig:LS_ST}
\end{figure}

As in the perturbed $XXZ$ chain, the $z$-component of the total spin is conserved, and therefore subspaces $\mathcal{\hat{S}}_{N}$ with a fixed number $N$ of spins up or down are used for the calculations.
To study the eigenstate IPR the diagonalized basis components are compared with the spin site basis and averaged over $10\%$ of the values in the center of the energy spectrum, obtaining $\bar{\xi}_{_{E}}$. Because of the statistical nature of the random variables $h_0,...,h_{L-1}$, $\bar{\xi}_{_{E}}$ is also averaged over several realizations but no new notation is added in order to prevent confusion.

{Figure~\ref{Fig:LS_FFT}(a)} shows the results for $\beta$ and $\bar{\xi}_{_{E}}$ for a spin chain of length $L=13$ with fixed number of spins $N=5$ ($D=1287$), averaged over 100 realizations. Transition from a Poisson NN level distribution to a WD one occurs at $h\approx0.5$ according to $\beta$. Similar results for $\bar{\xi}_{_{E}}$ were obtained for random perturbations with a Gaussian distribution in Ref.~\cite{Dukesz_2009}.

In {Fig.~\ref{Fig:LS_FFT} (b)} we show that both OTOC oscillations indicators $\bar{\xi}_{_{OTOC}}$ and $\bar{\sigma}_{_{OTOC}}^{-1}$, for a chain of length $L=9$ and $N=5$, follow a very similar behavior than of those shown by the quantum chaos indicators. The choice of a very small chain for the OTOC properties arises from the requirement of averaging many realizations, but it has been checked that a chain with the same parameters of those in the top panel of Fig.~\ref{Fig:LS_FFT} presents the same properties (data not shown).
We remark that it is not the scope of this work to try to identify the MBL transition present for large values of $h$ \cite{ALET2018}. However we note that the decline of $\bar{\xi}_{_{OTOC}}$ and  $\bar{\sigma}_{_{OTOC}}^{-1}$ with increasing $h$ points in the right direction to identify the MBL behavior.

Finally, we show in Fig.~\ref{Fig:LS_ST} the short-time growth of the OTOC $C_{zz}$ for this spin chain.
We also plot the short-time power-law relation that was obtained in Appendix~\ref{ap:short_time}.
We can see that Eq.~(\ref{short-hei}) describes very well this time regime. 
The random parameters $h_0,...h_{L-1}$, which cause the system to transition into chaotic regimes and then to the MBL, do not play a role in the short-time growth power-law. 


\section{Conclusions}
\label{sec:conclusions}
The OTOC is a quantity that has recently drawn attention since it has been suggested as a measure of quantum chaos, but also because of possible implications in studies of high energy physics, many-body localization, quantum information scrambling and thermalization.

After the so-called scrambling time the OTOC establishes around  a constant mean value. However there are fluctuations that remain and we had evidence that these fluctuations are strongly correlated to the dynamical properties of the system. We have proposed two quantities to assess these long-time fluctuations, the spectral IPR and the time variance, which we compared with well established chaos indicators like the Brody parameter and the localization of the eigenstates. We computed these quantities numerically for various one-body and many-body systems, which where known to undergo a  transition to chaos depending on one parameter.  

From our simulations we conclude that the fluctuations of the OTOC can be used to characterize a transition to chaos in quantum systems. For systems with a classical counterpart the spectral IPR can be directly related to the area of the regular islands in the phase space. For many-body spin systems the same qualitative behavior has been observed even though there is no classical counterpart. An important difference between the quantum chaos and the OTOC indicators emerges from the spin chain simulations. Unlike the latter, the former require to be calculated in eigenspaces without any remaining symmetry.
Besides, we also show that all the considered spin chains are slow scramblers due to the power-law growth of the OTOC for short times that does not depend on the regular to chaos transition.

Our results indicate that the main features of the dynamics can be extracted from the long time of the OTOC.  They suggest that this regime is feature rich and deserves more attention and study, in particular because the long-time regime matters in problems of current interest like quantum thermalization. They can also have implications in light of recent experimental advances.

\begin{acknowledgments}
We thank M. Saraceno and A. J. Roncaglia for fruitful discussions. We received funding from CONICET (Grant No. PIP 11220150100493CO.), ANCyPT (Grant No. PICT-2016-1056), UBACyT (Grant No. 20020170100234BA), the French National Research Agency (Project ANR- 14-CE36-0007-01), a bi-national collaboration project funded by CONICET and CNRS (PICS No. 06687) and the French-Argentinian Laboratoire International Associci\'e (LIA) ``LICoQ'' (funded by CNRS).  
\end{acknowledgments}

\appendix
\section{\label{ap:short_time}Short-time growth of the OTOC for spin chain models}

This appendix is devoted to depicting the basic steps to obtain the short-time behavior 
of the OTOC in the spin chain models.
The cornerstone of our calculations is the Hausdorff-Baker-Campbell (HBC) formula \cite{Oteo:1991kg}. The Heisenberg evolution of an operator $\hat{W}(t)$ can be expanded using the HBC formula as 
\begin{equation}\label{eq:HBC}
\hat{W}(t)= \sum_{n=0}^{\infty}\frac{\left(it\right)^{n}}{n!}\,\,\,\stackrel{\text{n times}}{[\hat{H},[\hat{H},...[\hat{H},}\hat{W}]]].
\end{equation}
 If $\hat{W}=\hat{\sigma}^{\mu}_l$ (the $\mu$-component of spin operator at site $l$), the HBC formula captures the spread of the operator over the spin sites and how it becomes more complex as time increases. Furthermore, direct replacement
 of Eq.~(\ref{eq:HBC}) in Eq.~(\ref{eq:OTOC_spin}) highlights the fact that the short-time growth is characterized by the smallest $n$ on which 
 \begin{equation}
     \stackrel{\text{n times}}{[\hat{H},[\hat{H},...[\hat{H},}\hat{\sigma}^{\mu}_0] ,\hat{\sigma}^{\nu}_l]]\neq 0,
 \end{equation}
 due to the time factor $t^n$ that weights the terms in the expansion. 
 We remark that this mechanism points out that the short-time growth is characterized by the general Hamiltonian structure of the system and not by the regular to chaotic regimes observed in the studied spin chains.

We consider first the Heisenberg spin chain with random magnetic field. Using Eq. \ref{eq:HBC}, the Heisenberg evolution of the spin operator $\hat{\sigma}_{0}^z$ is obtained, 
\begin{equation}\label{eq:HBC_App}
\hat{\sigma}_0^z (t) = \hat{\sigma}_0^z+it[\hat{H},\hat{\sigma}_0^z]+\frac{\left(it\right)^{2}}{2}[\hat{H},[\hat{H},\hat{\sigma}_0^z]+...\end{equation}
It is straightforward to show that the first and second  order terms of Eq. \ref{eq:HBC_App} result, respectively, in
\begin{equation}\label{eq:appendix1}
[\hat{H},\hat{\sigma}_0^z] =\frac{i}{2}\left(\hat{\sigma}_{0}^{x}\hat{\sigma}_{1}^{y}-\hat{\sigma}_{0}^{y}\hat{\sigma}_{1}^{x}\right),
\end{equation}

\begin{multline}
\small{
    [\hat{H},[\hat{H},\hat{\sigma}_{0}^{z}]]=\frac{\hat{\sigma}_{0}^{z}-\hat{\sigma}_{1}^{z}}{2}-\frac{h_{0}-h_{1}}{2}\left(\hat{\sigma}_{0}^{x}\hat{\sigma}_{1}^{x}+\hat{\sigma}_{0}^{y}\hat{\sigma}_{1}^{y}\right)}
    \\\small{-\frac{1}{4}\left(\hat{\sigma}_{0}^{x}\hat{\sigma}_{1}^{z}\hat{\sigma}_{2}^{x}+\hat{\sigma}_{0}^{y}\hat{\sigma}_{1}^{z}\hat{\sigma}_{2}^{y}-\hat{\sigma}_{0}^{x}\hat{\sigma}_{1}^{x}\hat{\sigma}_{2}^{z}-\hat{\sigma}_{0}^{y}\hat{\sigma}_{1}^{y}\hat{\sigma}_{2}^{z}\right)}.
\end{multline}

From these expressions, we see that the $l$-th order HBC term includes spin operators up to site $l$. Then, replacing Eq.~(\ref{eq:HBC_App}) in the OTOC expression of Eq.~(\ref{eq:OTOC_spin}), we obtain
\begin{equation}
C_{zz}\left(l,t\right)=\frac{1}{2}\left\langle \left|\left[\left(\hat{\sigma}_0^z+it[\hat{H},\hat{\sigma}_0^z]+...\right),\hat{\sigma}_{l}^{z}\right]\right|^{2}\right\rangle,
\end{equation}
from which is clear that the first nonzero term is of order $l$. This term dominates in the short-time regime over the following ones, due to the $t^l$ weights present in the HBC formula. Therefore, the short-time growth is given by 
\begin{equation}
C_{zz}\left(l,t\right)\approx\frac{1}{2}\frac{t^{2l}}{\left(l!\right)^{2}},
\label{short-hei}
\end{equation}
for $l\geq1$. The procedure for the other two spin chains is similar. For the Ising model with tilted magnetic field the first two relevant terms of the HBC expansion are, respectively, 
\begin{equation}
    [\hat{H},\hat{\sigma}_{0}^{z}] = -i\frac{B}{2}\sin\left(\theta \right)\hat{\sigma}_0^y,
\end{equation}

\begin{multline}
    [\hat{H},[\hat{H},[\hat{H},\hat{\sigma}_{0}^{z}]]]=-iB^{3}sin(\theta)\left(\frac{J^{2}}{4B^{2}}+1\right)\hat{\sigma}_{0}^{y}
    \\+iJB^{2}\left(\frac{sin^{2}(\theta)}{2}\hat{\sigma}_{0}^{x}\hat{\sigma}_{1}^{y}-sin(\theta)cos(\theta)\hat{\sigma}_{0}^{y}\hat{\sigma}_{1}^{z}\right).
\end{multline}
Then, it is expected that
\begin{equation}
C_{zz}\left(l,t\right)\approx \frac{1}{2} \left(B\sin\left(\theta\right)\right)^{2(2l+1)}\frac{t^{2(2l+1)}}{\left((2l+1)!\right)^2}
\label{stilt}.
\end{equation}

\begin{figure}[h]
\centering
\includegraphics[width=0.9\linewidth]{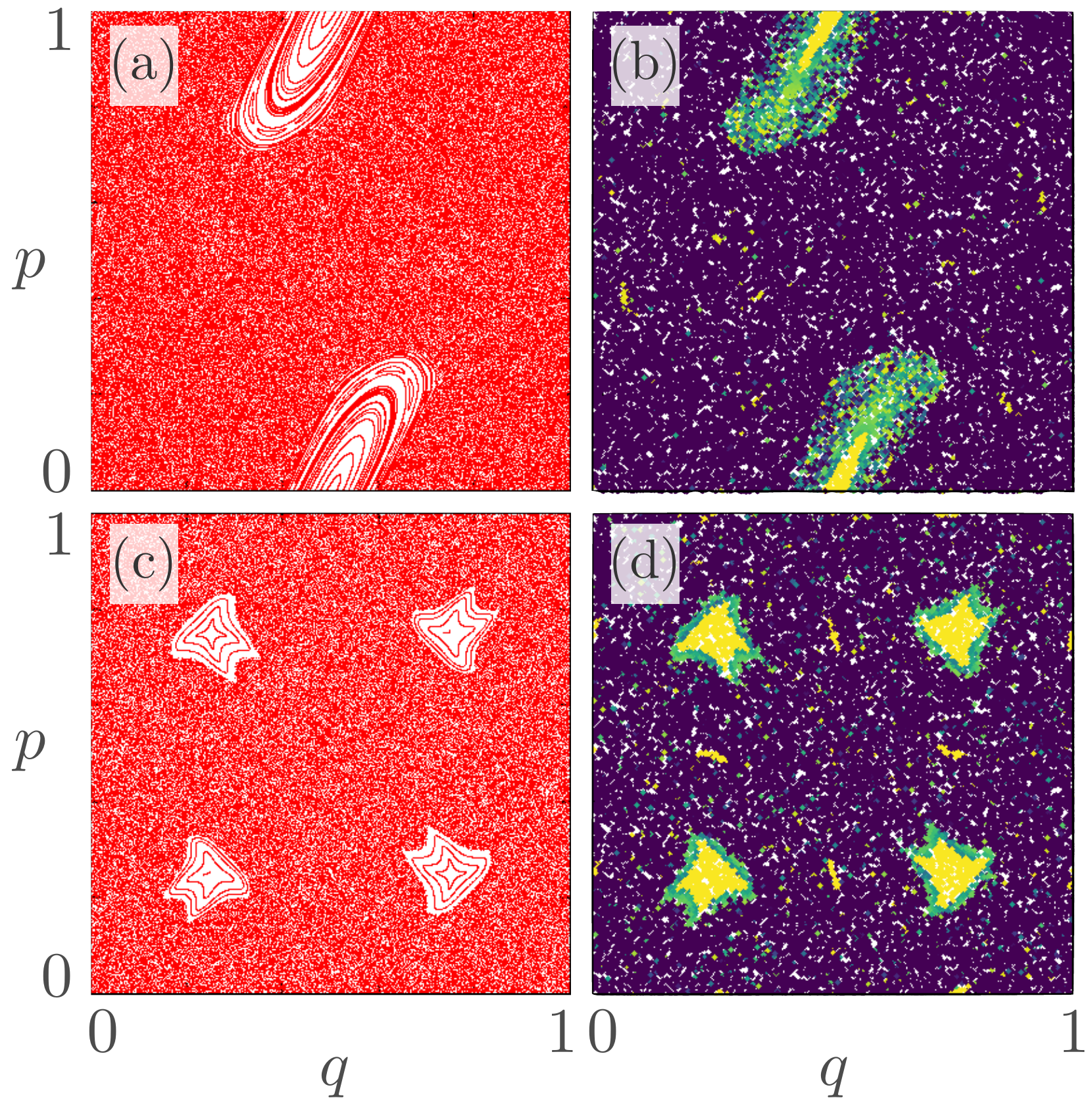} 
\caption{{(a),(c)}: classical phase portrait for the standard map {(a)} with $K=3.0$ and the Harper map {(c)} with $K=0.5$. {(b),(d)}: evolution in time for $30000$ randomly chosen initial conditions and the color represents $t_{max}$. Light colors stand for small evolution times, where the point has returned to the vicinity of the initial position before the $t_{max}$. The darkest color stands for the cases where the point has not returned to the neighborhood of the initial condition within a time $t_{max}$. The ratio (\ref{app:ratio}) is $r_{ch}=0.882$ for the standard map {(b)} and $r_{ch}=0.866$ for the Harper map {(d)}, meaning that $86 $--$ 88\,\%$ (approx.) of the area of the phase-space is chaotic. }
\label{fig:classical}   
\end{figure}
The  presence of the NNN interaction in the perturbed $XXZ$ model plays a significant role in the short-time 
behavior of the OTOC. 
Following the previous analysis, it is easy to see that the short-time growth power law is characterized by
\begin{equation}
C_{zz}\left(l,t\right)\approx \begin{cases} \frac{1}{2}\frac{t^{2l}}{\left(l!\right)^{2}} & \mbox{if } l=1 \,\vee\, \lambda=0,l>1 \\ \frac{1}{2}\frac{\lambda^{2\left(l-1\right)}t^{2\left(l-1\right)}}{\left[\left(l-1\right)!\right]^{2}} & \mbox{if }\lambda\neq 0,l\geq2 \end{cases}  
\label{shorttime1}
\end{equation}
We note that the result depends on whether the perturbation (proportional to $\lambda$) is present or not. In the latter case, the short-time behavior of the  Heisenberg spin chain with random magnetic field is recovered.

\section{Metropolis sampling of the  chaotic area in the phase space for classical maps}
\label{app:metro}
In this appendix we establish a classical measure in order to gauge the chaotic fraction of the phase space of a map. We use a Metropolis-like approach in order to measure the area of the chaotic region in  phase-space. We randomly choose a number $n_{tot}$ of initial conditions $\{q_i,p_i\}_{i=1}^{n_{tot}}$, large enough to sufficiently sample all the unit square (typically of the order $2\times 10^4$). We evolve these points with the classical map up to a chosen number of iterations, $t_{max}$. In cases in which the evolution drives the point within a predefined distance $\delta$ from the initial conditions, the evolution stops and the final time is recorded. If $t_{max}$ is not too large -- this point is crucial -- most of the points in the chaotic region will arrive back at the neighborhood of the initial point, and thus the final time recorded will be $t_{max}$. We then define the ratio
\begin{equation}
    r_{ch}=\frac{n_{t_{max}}}{n_{tot}},
\label{app:ratio}
\end{equation}
where $n_{t_{max}}$ is the number of initial conditions which have not returned after $t_{max}$ iterations.
The complement $r_{reg} \equiv 1-r_{ch}$ gives an estimation of the area of points inside regular islands. We illustrate in Fig.~\ref{fig:classical} the appropriateness of our classical measure through color-coded examples for the standard and the Harper maps. It can be seen that the darker color in the right columns for both cases approximates well the area of the chaotic region that can be identified on the phase portraits depicted in the left columns.

{In the calculations presented in the main text $r_{ch}$ was averaged over $20$ contiguous $t_{max}$ to  smoother out the effect of the choice of $t_{max}$ and $\delta$.}

\begin{thebibliography}{96}
\expandafter\ifx\csname natexlab\endcsname\relax\def\natexlab#1{#1}\fi
\expandafter\ifx\csname bibnamefont\endcsname\relax
  \def\bibnamefont#1{#1}\fi
\expandafter\ifx\csname bibfnamefont\endcsname\relax
  \def\bibfnamefont#1{#1}\fi
\expandafter\ifx\csname citenamefont\endcsname\relax
  \def\citenamefont#1{#1}\fi
\expandafter\ifx\csname url\endcsname\relax
  \def\url#1{\texttt{#1}}\fi
\expandafter\ifx\csname urlprefix\endcsname\relax\def\urlprefix{URL }\fi
\providecommand{\bibinfo}[2]{#2}
\providecommand{\eprint}[2][]{\url{#2}}

\bibitem[{\citenamefont{Einstein and Engel}(1987)}]{einstein}
\bibinfo{author}{\bibfnamefont{A.}~\bibnamefont{Einstein}} \bibnamefont{and}
  \bibinfo{author}{\bibfnamefont{A.}~\bibnamefont{Engel}},
  \emph{\bibinfo{title}{The Collected Papers of Albert Einstein}},
  vol.~\bibinfo{volume}{7} (\bibinfo{publisher}{Princeton University Press, Princeton},
  \bibinfo{year}{1987}).

\bibitem[{\citenamefont{Stone}(2005)}]{stone}
\bibinfo{author}{\bibfnamefont{A.~D.} \bibnamefont{Stone}},
  \bibinfo{journal}{Physics Today} \textbf{\bibinfo{volume}{58}},
  \bibinfo{pages}{37} (\bibinfo{year}{2005}).

\bibitem[{\citenamefont{Gutzwiller}(1990)}]{gutzwiller}
\bibinfo{author}{\bibfnamefont{M.~C.} \bibnamefont{Gutzwiller}},
  \emph{\bibinfo{title}{Chaos in Classical and Quantum Mechanics}}
  (\bibinfo{publisher}{Springer-Verlag, New York}, \bibinfo{year}{1990}).

\bibitem[{\citenamefont{Bohigas et~al.}(1984)\citenamefont{Bohigas,
  J.~Giannoni, and Schmit}}]{Bohigas}
\bibinfo{author}{\bibfnamefont{O.}~\bibnamefont{Bohigas}},
  \bibinfo{author}{\bibfnamefont{M.}~\bibnamefont{J.~Giannoni}},
  \bibnamefont{and} \bibinfo{author}{\bibfnamefont{C.}~\bibnamefont{Schmit}},
  \bibinfo{journal}{Phys. Rev. Lett.} \textbf{\bibinfo{volume}{52}},
  \bibinfo{pages}{1} (\bibinfo{year}{1984}).

\bibitem[{\citenamefont{Lewis-Swan et~al.}(2019)\citenamefont{Lewis-Swan,
  Safavi-Naini, Bollinger, and Rey}}]{LewisSwan:2019gj}
\bibinfo{author}{\bibfnamefont{R.~J.} \bibnamefont{Lewis-Swan}},
  \bibinfo{author}{\bibfnamefont{A.}~\bibnamefont{Safavi-Naini}},
  \bibinfo{author}{\bibfnamefont{J.~J.} \bibnamefont{Bollinger}},
  \bibnamefont{and} \bibinfo{author}{\bibfnamefont{A.~M.} \bibnamefont{Rey}},
  \bibinfo{journal}{Nature Communications} \textbf{\bibinfo{volume}{10}},
  \bibinfo{pages}{1581} (\bibinfo{year}{2019}).

\bibitem[{\citenamefont{Hess et~al.}(2017)\citenamefont{Hess, Becker, Kaplan,
  Kyprianidis, Lee, Neyenhuis, Pagano, Richerme, Senko, Smith
  et~al.}}]{Monroe1}
\bibinfo{author}{\bibfnamefont{P.~W.} \bibnamefont{Hess}},
  \bibinfo{author}{\bibfnamefont{P.}~\bibnamefont{Becker}},
  \bibinfo{author}{\bibfnamefont{H.~B.} \bibnamefont{Kaplan}},
  \bibinfo{author}{\bibfnamefont{A.}~\bibnamefont{Kyprianidis}},
  \bibinfo{author}{\bibfnamefont{A.~C.} \bibnamefont{Lee}},
  \bibinfo{author}{\bibfnamefont{B.}~\bibnamefont{Neyenhuis}},
  \bibinfo{author}{\bibfnamefont{G.}~\bibnamefont{Pagano}},
  \bibinfo{author}{\bibfnamefont{P.}~\bibnamefont{Richerme}},
  \bibinfo{author}{\bibfnamefont{C.}~\bibnamefont{Senko}},
  \bibinfo{author}{\bibfnamefont{J.}~\bibnamefont{Smith}},
  \bibnamefont{et~al.}, \bibinfo{journal}{Phil. Trans. R. Soc. A}
  \textbf{\bibinfo{volume}{375}}, \bibinfo{pages}{0107} (\bibinfo{year}{2017}).

\bibitem[{\citenamefont{Kaufman et~al.}(2016)\citenamefont{Kaufman, Tai, Lukin,
  Rispoli, Schittko, Preiss, and Greiner}}]{Kaufman794}
\bibinfo{author}{\bibfnamefont{A.~M.} \bibnamefont{Kaufman}},
  \bibinfo{author}{\bibfnamefont{M.~E.} \bibnamefont{Tai}},
  \bibinfo{author}{\bibfnamefont{A.}~\bibnamefont{Lukin}},
  \bibinfo{author}{\bibfnamefont{M.}~\bibnamefont{Rispoli}},
  \bibinfo{author}{\bibfnamefont{R.}~\bibnamefont{Schittko}},
  \bibinfo{author}{\bibfnamefont{P.~M.} \bibnamefont{Preiss}},
  \bibnamefont{and} \bibinfo{author}{\bibfnamefont{M.}~\bibnamefont{Greiner}},
  \bibinfo{journal}{Science}
  \textbf{\bibinfo{volume}{353}}, \bibinfo{pages}{794} (\bibinfo{year}{2016}).

\bibitem[{\citenamefont{Schreiber et~al.}(2015)\citenamefont{Schreiber,
  Hodgman, Bordia, L{\"u}schen, Fischer, Vosk, Altman, Schneider, and
  Bloch}}]{Schreiber842}
\bibinfo{author}{\bibfnamefont{M.}~\bibnamefont{Schreiber}},
  \bibinfo{author}{\bibfnamefont{S.~S.} \bibnamefont{Hodgman}},
  \bibinfo{author}{\bibfnamefont{P.}~\bibnamefont{Bordia}},
  \bibinfo{author}{\bibfnamefont{H.~P.} \bibnamefont{L{\"u}schen}},
  \bibinfo{author}{\bibfnamefont{M.~H.} \bibnamefont{Fischer}},
  \bibinfo{author}{\bibfnamefont{R.}~\bibnamefont{Vosk}},
  \bibinfo{author}{\bibfnamefont{E.}~\bibnamefont{Altman}},
  \bibinfo{author}{\bibfnamefont{U.}~\bibnamefont{Schneider}},
  \bibnamefont{and} \bibinfo{author}{\bibfnamefont{I.}~\bibnamefont{Bloch}},
    \bibinfo{journal}{Science}
  \textbf{\bibinfo{volume}{349}}, \bibinfo{pages}{842} (\bibinfo{year}{2015}).

\bibitem[{\citenamefont{Li et~al.}(2017)\citenamefont{Li, Fan, Wang, Ye, Zeng,
  Zhai, Peng, and Du}}]{Li:2017ic}
\bibinfo{author}{\bibfnamefont{J.}~\bibnamefont{Li}},
  \bibinfo{author}{\bibfnamefont{R.}~\bibnamefont{Fan}},
  \bibinfo{author}{\bibfnamefont{H.}~\bibnamefont{Wang}},
  \bibinfo{author}{\bibfnamefont{B.}~\bibnamefont{Ye}},
  \bibinfo{author}{\bibfnamefont{B.}~\bibnamefont{Zeng}},
  \bibinfo{author}{\bibfnamefont{H.}~\bibnamefont{Zhai}},
  \bibinfo{author}{\bibfnamefont{X.}~\bibnamefont{Peng}}, \bibnamefont{and}
  \bibinfo{author}{\bibfnamefont{J.}~\bibnamefont{Du}}, \bibinfo{journal}{Phys.
  Rev. X} \textbf{\bibinfo{volume}{7}}, \bibinfo{pages}{031011}
  (\bibinfo{year}{2017}).

\bibitem[{\citenamefont{{Larkin} and
  {Ovchinnikov}}(1969)}]{1969JETP...28.1200L}
\bibinfo{author}{\bibfnamefont{A.~I.} \bibnamefont{{Larkin}}} \bibnamefont{and}
  \bibinfo{author}{\bibfnamefont{Y.~N.} \bibnamefont{{Ovchinnikov}}},
  \bibinfo{journal}{Soviet Journal of Experimental and Theoretical Physics}
  \textbf{\bibinfo{volume}{28}}, \bibinfo{pages}{1200} (\bibinfo{year}{1969}).

\bibitem[{\citenamefont{Maldacena et~al.}(2015)\citenamefont{Maldacena,
  H.~Shenker, and Stanford}}]{Bound}
\bibinfo{author}{\bibfnamefont{J.}~\bibnamefont{Maldacena}},
  \bibinfo{author}{\bibfnamefont{S.}~\bibnamefont{H.~Shenker}},
  \bibnamefont{and} \bibinfo{author}{\bibfnamefont{D.}~\bibnamefont{Stanford}},
  \bibinfo{journal}{J. High Energ. Phys. (2016) 2016: 106. }

\bibitem[{\citenamefont{Swingle et~al.}(2016)\citenamefont{Swingle, Bentsen,
  Schleier-Smith, and Hayden}}]{PhysRevA.94.040302}
\bibinfo{author}{\bibfnamefont{B.}~\bibnamefont{Swingle}},
  \bibinfo{author}{\bibfnamefont{G.}~\bibnamefont{Bentsen}},
  \bibinfo{author}{\bibfnamefont{M.}~\bibnamefont{Schleier-Smith}},
  \bibnamefont{and} \bibinfo{author}{\bibfnamefont{P.}~\bibnamefont{Hayden}},
  \bibinfo{journal}{Phys. Rev. A} \textbf{\bibinfo{volume}{94}},
  \bibinfo{pages}{040302} (\bibinfo{year}{2016}).

\bibitem[{\citenamefont{Riddell and S{\o}rensen}(2019)}]{Riddell:2019ed}
\bibinfo{author}{\bibfnamefont{J.}~\bibnamefont{Riddell}} \bibnamefont{and}
  \bibinfo{author}{\bibfnamefont{E.~S.} \bibnamefont{S{\o}rensen}},
  \bibinfo{journal}{Phys. Rev. B} \textbf{\bibinfo{volume}{99}},
  \bibinfo{pages}{054205} (\bibinfo{year}{2019}).

\bibitem[{\citenamefont{Landsman et~al.}(2019)\citenamefont{Landsman, Figgatt,
  Schuster, Linke, Yoshida, Yao, and Monroe}}]{Landsman:2019ke}
\bibinfo{author}{\bibfnamefont{K.~A.} \bibnamefont{Landsman}},
  \bibinfo{author}{\bibfnamefont{C.}~\bibnamefont{Figgatt}},
  \bibinfo{author}{\bibfnamefont{T.}~\bibnamefont{Schuster}},
  \bibinfo{author}{\bibfnamefont{N.~M.} \bibnamefont{Linke}},
  \bibinfo{author}{\bibfnamefont{B.}~\bibnamefont{Yoshida}},
  \bibinfo{author}{\bibfnamefont{N.~Y.} \bibnamefont{Yao}}, \bibnamefont{and}
  \bibinfo{author}{\bibfnamefont{C.}~\bibnamefont{Monroe}},
  \bibinfo{journal}{Nature (London)} \textbf{\bibinfo{volume}{567}}, \bibinfo{pages}{61}
  (\bibinfo{year}{2019}).

\bibitem[{\citenamefont{Swingle and Chowdhury}(2017)}]{SwingleSlow}
\bibinfo{author}{\bibfnamefont{B.}~\bibnamefont{Swingle}} \bibnamefont{and}
  \bibinfo{author}{\bibfnamefont{D.}~\bibnamefont{Chowdhury}},
  \bibinfo{journal}{Phys. Rev. B} \textbf{\bibinfo{volume}{95}},
  \bibinfo{pages}{060201} (\bibinfo{year}{2017}).

\bibitem[{\citenamefont{Chen et~al.}(2017)\citenamefont{Chen, Zhou, Huse, and
  Fradkin}}]{chen2017out}
\bibinfo{author}{\bibfnamefont{X.}~\bibnamefont{Chen}},
  \bibinfo{author}{\bibfnamefont{T.}~\bibnamefont{Zhou}},
  \bibinfo{author}{\bibfnamefont{D.~A.} \bibnamefont{Huse}}, \bibnamefont{and}
  \bibinfo{author}{\bibfnamefont{E.}~\bibnamefont{Fradkin}},
  \bibinfo{journal}{Annalen der Physik} \textbf{\bibinfo{volume}{529}},
  \bibinfo{pages}{1600332} (\bibinfo{year}{2017}).

\bibitem[{\citenamefont{Slagle et~al.}(2017)\citenamefont{Slagle, Bi, You, and
  Xu}}]{SlaglePRB2017}
\bibinfo{author}{\bibfnamefont{K.}~\bibnamefont{Slagle}},
  \bibinfo{author}{\bibfnamefont{Z.}~\bibnamefont{Bi}},
  \bibinfo{author}{\bibfnamefont{Y.-Z.} \bibnamefont{You}}, \bibnamefont{and}
  \bibinfo{author}{\bibfnamefont{C.}~\bibnamefont{Xu}}, \bibinfo{journal}{Phys.
  Rev. B} \textbf{\bibinfo{volume}{95}}, \bibinfo{pages}{165136}
  (\bibinfo{year}{2017}).

\bibitem[{\citenamefont{Luitz and Bar~Lev}(2017)}]{Luitz2017}
\bibinfo{author}{\bibfnamefont{D.~J.} \bibnamefont{Luitz}} \bibnamefont{and}
  \bibinfo{author}{\bibfnamefont{Y.}~\bibnamefont{Bar~Lev}},
  \bibinfo{journal}{Phys. Rev. B} \textbf{\bibinfo{volume}{96}},
  \bibinfo{pages}{020406} (\bibinfo{year}{2017}).

\bibitem[{\citenamefont{Subhayan~Sahu}(2018)}]{Sahu}
\bibinfo{author}{\bibfnamefont{B.~S.} \bibnamefont{Subhayan~Sahu} and
  \bibfnamefont{Shenglong~Xu}}, \bibinfo{journal}{arXiv:1807.06086}
  (\bibinfo{year}{2018}).

\bibitem[{\citenamefont{G{\"a}rttner et~al.}(2017)\citenamefont{G{\"a}rttner,
  Bohnet, Safavi-Naini, Wall, Bollinger, and Rey}}]{garttner2017measuring}
\bibinfo{author}{\bibfnamefont{M.}~\bibnamefont{G{\"a}rttner}},
  \bibinfo{author}{\bibfnamefont{J.~G.} \bibnamefont{Bohnet}},
  \bibinfo{author}{\bibfnamefont{A.}~\bibnamefont{Safavi-Naini}},
  \bibinfo{author}{\bibfnamefont{M.~L.} \bibnamefont{Wall}},
  \bibinfo{author}{\bibfnamefont{J.~J.} \bibnamefont{Bollinger}},
  \bibnamefont{and} \bibinfo{author}{\bibfnamefont{A.~M.} \bibnamefont{Rey}},
  \bibinfo{journal}{Nature Physics} \textbf{\bibinfo{volume}{13}},
  \bibinfo{pages}{781} (\bibinfo{year}{2017}).

\bibitem[{\citenamefont{Wei et~al.}(2018)\citenamefont{Wei, Peng, Shtanko,
  Marvian, Lloyd, Ramanathan, and Cappellaro}}]{cappellaro2018}
\bibinfo{author}{\bibfnamefont{K.~X.} \bibnamefont{Wei}},
  \bibinfo{author}{\bibfnamefont{P.}~\bibnamefont{Peng}},
  \bibinfo{author}{\bibfnamefont{O.}~\bibnamefont{Shtanko}},
  \bibinfo{author}{\bibfnamefont{I.}~\bibnamefont{Marvian}},
  \bibinfo{author}{\bibfnamefont{S.}~\bibnamefont{Lloyd}},
  \bibinfo{author}{\bibfnamefont{C.}~\bibnamefont{Ramanathan}},
  \bibnamefont{and}
  \bibinfo{author}{\bibfnamefont{P.}~\bibnamefont{Cappellaro}},
  \bibinfo{journal}{arXiv:1812.04776}  (\bibinfo{year}{2018}).

\bibitem[{\citenamefont{Sánchez et~al.}(2019)\citenamefont{Sánchez, Chattah,
  Wei, Buljubasich, Cappellaro, and Pastawski}}]{pastawski2019}
\bibinfo{author}{\bibfnamefont{C.~M.} \bibnamefont{Sánchez}},
  \bibinfo{author}{\bibfnamefont{A.~K.} \bibnamefont{Chattah}},
  \bibinfo{author}{\bibfnamefont{K.~X.} \bibnamefont{Wei}},
  \bibinfo{author}{\bibfnamefont{L.}~\bibnamefont{Buljubasich}},
  \bibinfo{author}{\bibfnamefont{P.}~\bibnamefont{Cappellaro}},
  \bibnamefont{and} \bibinfo{author}{\bibfnamefont{H.~M.}
  \bibnamefont{Pastawski}}, \bibinfo{journal}{arXiv:1902.06628}
  (\bibinfo{year}{2019}).

\bibitem[{\citenamefont{Sachdev and Ye}(1993)}]{SachdevYe}
\bibinfo{author}{\bibfnamefont{S.}~\bibnamefont{Sachdev}} \bibnamefont{and}
  \bibinfo{author}{\bibfnamefont{J.}~\bibnamefont{Ye}}, \bibinfo{journal}{Phys.
  Rev. Lett.} \textbf{\bibinfo{volume}{70}}, \bibinfo{pages}{3339}
  (\bibinfo{year}{1993}).

\bibitem[{\citenamefont{Kitaev}(2015)}]{Kitaev}
\bibinfo{author}{\bibfnamefont{A.}~\bibnamefont{Kitaev}},
 \bibinfo{title}{A simple model of quantum holography}, 
 \bibinfo{journal}{talk given at
KITP Program: Entanglement in Strongly-Correlated Quantum Matter, http://online.kitp.ucsb.edu/online/entangled15/kitaev/ (April 7, 2015), http://online.kitp.ucsb.edu/online/entangled15/ kitaev2/ (May 27, 2015).}
  
 

\bibitem[{\citenamefont{H.~Shenker and Stanford}(2013)}]{AdS}
\bibinfo{author}{\bibfnamefont{S.}~\bibnamefont{H.~Shenker}} \bibnamefont{and}
  \bibinfo{author}{\bibfnamefont{D.}~\bibnamefont{Stanford}},
  \bibinfo{journal}{ J. High Energ. Phys. (2014) 2014: 67}.

\bibitem[{\citenamefont{Ch{\'a}vez-Carlos
  et~al.}(2019)\citenamefont{Ch{\'a}vez-Carlos, L{\'o}pez-del Carpio,
  Bastarrachea-Magnani, Str{\'a}nsk{\'{y}}, Lerma-Hern{\'a}ndez, Santos, and
  Hirsch}}]{ChavezCarlos:2019bn}
\bibinfo{author}{\bibfnamefont{J.}~\bibnamefont{Ch{\'a}vez-Carlos}},
  \bibinfo{author}{\bibfnamefont{B.}~\bibnamefont{L{\'o}pez-del Carpio}},
  \bibinfo{author}{\bibfnamefont{M.~A.} \bibnamefont{Bastarrachea-Magnani}},
  \bibinfo{author}{\bibfnamefont{P.}~\bibnamefont{Str{\'a}nsk{\'{y}}}},
  \bibinfo{author}{\bibfnamefont{S.}~\bibnamefont{Lerma-Hern{\'a}ndez}},
  \bibinfo{author}{\bibfnamefont{L.~F.} \bibnamefont{Santos}},
  \bibnamefont{and} \bibinfo{author}{\bibfnamefont{J.~G.}
  \bibnamefont{Hirsch}}, \bibinfo{journal}{Phys. Rev. Lett.}
  \textbf{\bibinfo{volume}{122}}, \bibinfo{pages}{024101}
  (\bibinfo{year}{2019}).

\bibitem[{\citenamefont{Roberts et~al.}(2018)\citenamefont{Roberts, Stanford,
  and Streicher}}]{Roberts:2018ck}
\bibinfo{author}{\bibfnamefont{D.~A.} \bibnamefont{Roberts}},
  \bibinfo{author}{\bibfnamefont{D.}~\bibnamefont{Stanford}}, \bibnamefont{and}
  \bibinfo{author}{\bibfnamefont{A.}~\bibnamefont{Streicher}},
  \bibinfo{journal}{Journal of High Energy Physics}
  \textbf{\bibinfo{volume}{2018}}, \bibinfo{pages}{122} (\bibinfo{year}{2018}).

\bibitem[{\citenamefont{Maldacena and Stanford}(2016)}]{Maldacena:2016hq}
\bibinfo{author}{\bibfnamefont{J.}~\bibnamefont{Maldacena}} \bibnamefont{and}
  \bibinfo{author}{\bibfnamefont{D.}~\bibnamefont{Stanford}},
  \bibinfo{journal}{Phys. Rev. D} \textbf{\bibinfo{volume}{94}},
  \bibinfo{pages}{106002} (\bibinfo{year}{2016}).

\bibitem[{\citenamefont{Sekino and Susskind}(2008)}]{Sekino_2008}
\bibinfo{author}{\bibfnamefont{Y.}~\bibnamefont{Sekino}} \bibnamefont{and}
  \bibinfo{author}{\bibfnamefont{L.}~\bibnamefont{Susskind}},
  \bibinfo{journal}{J. H. Ener. Phys.} \textbf{\bibinfo{volume}{2008}},
  \bibinfo{pages}{065} (\bibinfo{year}{2008}).

\bibitem[{\citenamefont{Garc{\'\i}a-Mata
  et~al.}(2018)\citenamefont{Garc{\'\i}a-Mata, Saraceno, Jalabert, Roncaglia,
  and Wisniacki}}]{GarciaMata:2018kz}
\bibinfo{author}{\bibfnamefont{I.}~\bibnamefont{Garc{\'\i}a-Mata}},
  \bibinfo{author}{\bibfnamefont{M.}~\bibnamefont{Saraceno}},
  \bibinfo{author}{\bibfnamefont{R.~A.} \bibnamefont{Jalabert}},
  \bibinfo{author}{\bibfnamefont{A.~J.} \bibnamefont{Roncaglia}},
  \bibnamefont{and} \bibinfo{author}{\bibfnamefont{D.~A.}
  \bibnamefont{Wisniacki}}, \bibinfo{journal}{Phys. Rev. Lett.}
  \textbf{\bibinfo{volume}{121}}, \bibinfo{pages}{210601}
  (\bibinfo{year}{2018}).

\bibitem[{\citenamefont{Jalabert et~al.}(2018)\citenamefont{Jalabert,
  Garc{\'\i}a-Mata, and Wisniacki}}]{Jalabert:2018da}
\bibinfo{author}{\bibfnamefont{R.~A.} \bibnamefont{Jalabert}},
  \bibinfo{author}{\bibfnamefont{I.}~\bibnamefont{Garc{\'\i}a-Mata}},
  \bibnamefont{and} \bibinfo{author}{\bibfnamefont{D.~A.}
  \bibnamefont{Wisniacki}}, \bibinfo{journal}{Phys. Rev. E}
  \textbf{\bibinfo{volume}{98}}, \bibinfo{pages}{062218}
  (\bibinfo{year}{2018}).

\bibitem[{\citenamefont{Rozenbaum et~al.}(2017)\citenamefont{Rozenbaum,
  Ganeshan, and Galitski}}]{Rozenbaum2017}
\bibinfo{author}{\bibfnamefont{E.~B.} \bibnamefont{Rozenbaum}},
  \bibinfo{author}{\bibfnamefont{S.}~\bibnamefont{Ganeshan}}, \bibnamefont{and}
  \bibinfo{author}{\bibfnamefont{V.}~\bibnamefont{Galitski}},
  \bibinfo{journal}{Phys. Rev. Lett.} \textbf{\bibinfo{volume}{118}},
  \bibinfo{pages}{086801} (\bibinfo{year}{2017}).

\bibitem[{\citenamefont{Chen and Zhou}(2018)}]{chen2018operator}
\bibinfo{author}{\bibfnamefont{X.}~\bibnamefont{Chen}} \bibnamefont{and}
  \bibinfo{author}{\bibfnamefont{T.}~\bibnamefont{Zhou}},
  \bibinfo{journal}{arXiv:1804.08655}  (\bibinfo{year}{2018}).

\bibitem[{\citenamefont{Lakshminarayan}(2019)}]{lakshminarayan2019out}
\bibinfo{author}{\bibfnamefont{A.}~\bibnamefont{Lakshminarayan}},
  \bibinfo{journal}{Phys. Rev. E} \textbf{\bibinfo{volume}{99}},
  \bibinfo{pages}{012201} (\bibinfo{year}{2019}).

\bibitem[{\citenamefont{D{\'o}ra and Moessner}(2017)}]{Dora:2017go}
\bibinfo{author}{\bibfnamefont{B.}~\bibnamefont{D{\'o}ra}} \bibnamefont{and}
  \bibinfo{author}{\bibfnamefont{R.}~\bibnamefont{Moessner}},
  \bibinfo{journal}{Phys. Rev. Lett.} \textbf{\bibinfo{volume}{119}},
  \bibinfo{pages}{026802} (\bibinfo{year}{2017}).

\bibitem[{\citenamefont{Kukuljan et~al.}(2017)\citenamefont{Kukuljan,
  Grozdanov, and Prosen}}]{ProsenWeak2017}
\bibinfo{author}{\bibfnamefont{I.}~\bibnamefont{Kukuljan}},
  \bibinfo{author}{\bibfnamefont{S.} \bibnamefont{Grozdanov}},
  \bibnamefont{and} \bibinfo{author}{\bibfnamefont{T.}
  \bibnamefont{Prosen}}, \bibinfo{journal}{Phys. Rev. B}
  \textbf{\bibinfo{volume}{96}}, \bibinfo{pages}{060301}
  (\bibinfo{year}{2017}).

\bibitem[{\citenamefont{Rammensee et~al.}(2018)\citenamefont{Rammensee, Urbina,
  and Richter}}]{rammensee2018many}
\bibinfo{author}{\bibfnamefont{J.}~\bibnamefont{Rammensee}},
  \bibinfo{author}{\bibfnamefont{J.~D.} \bibnamefont{Urbina}},
  \bibnamefont{and} \bibinfo{author}{\bibfnamefont{K.}~\bibnamefont{Richter}},
  \bibinfo{journal}{Phys. Rev. Lett.} \textbf{\bibinfo{volume}{121}},
  \bibinfo{pages}{124101} (\bibinfo{year}{2018}).

\bibitem[{\citenamefont{Hashimoto et~al.}(2017)\citenamefont{Hashimoto, Murata,
  and Yoshii}}]{hashimoto}
\bibinfo{author}{\bibfnamefont{K.}~\bibnamefont{Hashimoto}},
  \bibinfo{author}{\bibfnamefont{K.}~\bibnamefont{Murata}}, \bibnamefont{and}
  \bibinfo{author}{\bibfnamefont{R.}~\bibnamefont{Yoshii}},
  \bibinfo{journal}{J. High Energy Phys.} \textbf{\bibinfo{volume}{10}},
  \bibinfo{pages}{138} (\bibinfo{year}{2017}).

\bibitem[{\citenamefont{Cotler et~al.}(2018)\citenamefont{Cotler, Ding, and
  Penington}}]{cotler}
\bibinfo{author}{\bibfnamefont{J.~S.} \bibnamefont{Cotler}},
  \bibinfo{author}{\bibfnamefont{D.}~\bibnamefont{Ding}}, \bibnamefont{and}
  \bibinfo{author}{\bibfnamefont{G.~R.} \bibnamefont{Penington}},
  \bibinfo{journal}{Ann. Phys. (Amsterdam)} \textbf{\bibinfo{volume}{396}},
  \bibinfo{pages}{318} (\bibinfo{year}{2018}).

\bibitem[{\citenamefont{Prakash and
  Lakshminarayan}(2019)}]{prakash2019scrambling}
\bibinfo{author}{\bibfnamefont{R.}~\bibnamefont{Prakash}} \bibnamefont{and}
  \bibinfo{author}{\bibfnamefont{A.}~\bibnamefont{Lakshminarayan}},
  \bibinfo{journal}{arXiv:1904.06482}  (\bibinfo{year}{2019}).

\bibitem[{\citenamefont{Bergamasco et~al.}(2019)\citenamefont{Bergamasco,
  Carlo, and Rivas}}]{Bergamasco2019}
\bibinfo{author}{\bibfnamefont{P.~D.} \bibnamefont{Bergamasco}},
  \bibinfo{author}{\bibfnamefont{G.~G.} \bibnamefont{Carlo}}, \bibnamefont{and}
  \bibinfo{author}{\bibfnamefont{A.~M.~F.} \bibnamefont{Rivas}},
  \bibinfo{journal}{arXiv:1904.12830}  (\bibinfo{year}{2019}).

\bibitem[{\citenamefont{Hummel et~al.}(2018)\citenamefont{Hummel, Geiger,
  Urbina, and Richter}}]{hummel2018reversible}
\bibinfo{author}{\bibfnamefont{Q.}~\bibnamefont{Hummel}},
  \bibinfo{author}{\bibfnamefont{B.}~\bibnamefont{Geiger}},
  \bibinfo{author}{\bibfnamefont{J.~D.} \bibnamefont{Urbina}},
  \bibnamefont{and} \bibinfo{author}{\bibfnamefont{K.}~\bibnamefont{Richter}},
  \bibinfo{journal}{arXiv:1812.09237}  (\bibinfo{year}{2018}).

\bibitem[{\citenamefont{Chan et~al.}(2018)\citenamefont{Chan, De~Luca, and
  Chalker}}]{chan2018eigenstate}
\bibinfo{author}{\bibfnamefont{A.}~\bibnamefont{Chan}},
  \bibinfo{author}{\bibfnamefont{A.}~\bibnamefont{De~Luca}}, \bibnamefont{and}
  \bibinfo{author}{\bibfnamefont{J.}~\bibnamefont{Chalker}},
  \bibinfo{journal}{arXiv:1810.11014}  (\bibinfo{year}{2018}).

\bibitem[{\citenamefont{Khemani et~al.}(2018)\citenamefont{Khemani, Huse, and
  Nahum}}]{khemani2018velocity}
\bibinfo{author}{\bibfnamefont{V.}~\bibnamefont{Khemani}},
  \bibinfo{author}{\bibfnamefont{D.~A.} \bibnamefont{Huse}}, \bibnamefont{and}
  \bibinfo{author}{\bibfnamefont{A.}~\bibnamefont{Nahum}},
  \bibinfo{journal}{Physical Review B} \textbf{\bibinfo{volume}{98}},
  \bibinfo{pages}{144304} (\bibinfo{year}{2018}).

\bibitem[{\citenamefont{Borgonovi and Izrailev}(2019)}]{borgonovi2019emergence}
\bibinfo{author}{\bibfnamefont{F.}~\bibnamefont{Borgonovi}} \bibnamefont{and}
  \bibinfo{author}{\bibfnamefont{F.~M.} \bibnamefont{Izrailev}},
  \bibinfo{journal}{Physical Review E} \textbf{\bibinfo{volume}{99}},
  \bibinfo{pages}{012115} (\bibinfo{year}{2019}).

\bibitem[{\citenamefont{Bohrdt et~al.}(2017)\citenamefont{Bohrdt, Mendl,
  Endres, and Knap}}]{Bohrdt_2017}
\bibinfo{author}{\bibfnamefont{A.}~\bibnamefont{Bohrdt}},
  \bibinfo{author}{\bibfnamefont{C.~B.} \bibnamefont{Mendl}},
  \bibinfo{author}{\bibfnamefont{M.}~\bibnamefont{Endres}}, \bibnamefont{and}
  \bibinfo{author}{\bibfnamefont{M.}~\bibnamefont{Knap}}, \bibinfo{journal}{New
  J. Phys} \textbf{\bibinfo{volume}{19}}, \bibinfo{pages}{063001}
  (\bibinfo{year}{2017}).

\bibitem[{\citenamefont{Xu et~al.}(2019)\citenamefont{Xu, Li, Hsu, Swingle, and
  \mbox{Das Sarma}}}]{dassarma}
\bibinfo{author}{\bibfnamefont{S.}~\bibnamefont{Xu}},
  \bibinfo{author}{\bibfnamefont{X.}~\bibnamefont{Li}},
  \bibinfo{author}{\bibfnamefont{Y.-T.} \bibnamefont{Hsu}},
  \bibinfo{author}{\bibfnamefont{B.}~\bibnamefont{Swingle}}, \bibnamefont{and}
  \bibinfo{author}{\bibfnamefont{S.}~\bibnamefont{\mbox{Das Sarma}}},
  \bibinfo{journal}{arXiv:1902.07199}  (\bibinfo{year}{2019}).

\bibitem[{\citenamefont{Xu and Swingle}(2018{\natexlab{a}})}]{xu2018accessing}
\bibinfo{author}{\bibfnamefont{S.}~\bibnamefont{Xu}} \bibnamefont{and}
  \bibinfo{author}{\bibfnamefont{B.}~\bibnamefont{Swingle}},
  \bibinfo{journal}{arXiv:1802.00801}  (\bibinfo{year}{2018}{\natexlab{a}}).

\bibitem[{\citenamefont{Xu and Swingle}(2018{\natexlab{b}})}]{xu2018locality}
\bibinfo{author}{\bibfnamefont{S.}~\bibnamefont{Xu}} \bibnamefont{and}
  \bibinfo{author}{\bibfnamefont{B.}~\bibnamefont{Swingle}},
  \bibinfo{journal}{ Phys. Rev. X {\bf 9}, 031048 (2019)}.  

\bibitem[{\citenamefont{Brody et~al.}(1981)\citenamefont{Brody, Flores, French,
  Mello, Pandey, and Wong}}]{Brody}
\bibinfo{author}{\bibfnamefont{T.~A.} \bibnamefont{Brody}},
  \bibinfo{author}{\bibfnamefont{J.}~\bibnamefont{Flores}},
  \bibinfo{author}{\bibfnamefont{J.~B.} \bibnamefont{French}},
  \bibinfo{author}{\bibfnamefont{P.~A.} \bibnamefont{Mello}},
  \bibinfo{author}{\bibfnamefont{A.}~\bibnamefont{Pandey}}, \bibnamefont{and}
  \bibinfo{author}{\bibfnamefont{S.~S.~M.} \bibnamefont{Wong}},
  \bibinfo{journal}{Rev. Mod. Phys.} \textbf{\bibinfo{volume}{53}},
  \bibinfo{pages}{385} (\bibinfo{year}{1981}).

\bibitem[{\citenamefont{Berry and Tabor}(1977)}]{Level_statistics2}
\bibinfo{author}{\bibfnamefont{M.~V.} \bibnamefont{Berry}} \bibnamefont{and}
  \bibinfo{author}{\bibfnamefont{M.}~\bibnamefont{Tabor}},
  \bibinfo{journal}{R. Soc. London Proc. A}
  \textbf{\bibinfo{volume}{356}}, \bibinfo{pages}{375} (\bibinfo{year}{1977}).

\bibitem[{\citenamefont{Gubin and F.~Santos}(2011)}]{LS_ipr}
\bibinfo{author}{\bibfnamefont{A.}~\bibnamefont{Gubin}} \bibnamefont{and}
  \bibinfo{author}{\bibfnamefont{L.}~\bibnamefont{F.~Santos}},
  \bibinfo{journal}{Am. J. Phys.} \textbf{\bibinfo{volume}{80}},
  \bibinfo{pages}{246}
  (\bibinfo{year}{2012}).

\bibitem[{\citenamefont{Zelevinsky et~al.}(1996)\citenamefont{Zelevinsky,
  Brown, Frazier, and Horoi}}]{ZELEVINSKY199685}
\bibinfo{author}{\bibfnamefont{V.}~\bibnamefont{Zelevinsky}},
  \bibinfo{author}{\bibfnamefont{B.}~\bibnamefont{Brown}},
  \bibinfo{author}{\bibfnamefont{N.}~\bibnamefont{Frazier}}, \bibnamefont{and}
  \bibinfo{author}{\bibfnamefont{M.}~\bibnamefont{Horoi}},
  \bibinfo{journal}{Physics Reports} \textbf{\bibinfo{volume}{276}},
  \bibinfo{pages}{85 } (\bibinfo{year}{1996}). 
  
\bibitem[{\citenamefont{Stockmann}(1999)}]{stockmann_1999}
\bibinfo{author}{\bibfnamefont{H.-J.} \bibnamefont{Stockmann}},
  \emph{\bibinfo{title}{Quantum Chaos: An Introduction}}
  (\bibinfo{publisher}{Cambridge University Press}, \bibinfo{year}{1999}).

\bibitem[{\citenamefont{Berry and Robnik}(1984)}]{Berry_1984}
\bibinfo{author}{\bibfnamefont{M.~V.} \bibnamefont{Berry}} \bibnamefont{and}
  \bibinfo{author}{\bibfnamefont{M.}~\bibnamefont{Robnik}},
  \bibinfo{journal}{J. Phys. A}
  \textbf{\bibinfo{volume}{17}}, \bibinfo{pages}{2413} (\bibinfo{year}{1984}).

\bibitem[{\citenamefont{Prosen and Robnik}(1994)}]{Prosen_1994}
\bibinfo{author}{\bibfnamefont{T.}~\bibnamefont{Prosen}} \bibnamefont{and}
  \bibinfo{author}{\bibfnamefont{M.}~\bibnamefont{Robnik}},
  \bibinfo{journal}{ J. Phys. A: Math. Gen.}
  \textbf{\bibinfo{volume}{27}}, \bibinfo{pages}{8059} (\bibinfo{year}{1994}).

\bibitem[{\citenamefont{Oganesyan and Huse}(2007)}]{Oganesyan_2007}
\bibinfo{author}{\bibfnamefont{V.}~\bibnamefont{Oganesyan}} \bibnamefont{and}
  \bibinfo{author}{\bibfnamefont{D.~A.} \bibnamefont{Huse}},
  \bibinfo{journal}{Phys. Rev. B} \textbf{\bibinfo{volume}{75}},
  \bibinfo{pages}{155111} (\bibinfo{year}{2007}).

\bibitem[{\citenamefont{Kudo and Deguchi}(2018)}]{KazueTetsuo_2018}
\bibinfo{author}{\bibfnamefont{K.}~\bibnamefont{Kudo}} \bibnamefont{and}
  \bibinfo{author}{\bibfnamefont{T.}~\bibnamefont{Deguchi}},
  \bibinfo{journal}{Phys. Rev. B} \textbf{\bibinfo{volume}{97}},
  \bibinfo{pages}{220201} (\bibinfo{year}{2018}).

\bibitem[{\citenamefont{Atas et~al.}(2013)\citenamefont{Atas, Bogomolny,
  Giraud, and Roux}}]{Atas_2013}
\bibinfo{author}{\bibfnamefont{Y.~Y.} \bibnamefont{Atas}},
  \bibinfo{author}{\bibfnamefont{E.}~\bibnamefont{Bogomolny}},
  \bibinfo{author}{\bibfnamefont{O.}~\bibnamefont{Giraud}}, \bibnamefont{and}
  \bibinfo{author}{\bibfnamefont{G.}~\bibnamefont{Roux}},
  \bibinfo{journal}{Phys. Rev. Lett.} \textbf{\bibinfo{volume}{110}},
  \bibinfo{pages}{084101} (\bibinfo{year}{2013}).

\bibitem[{\citenamefont{Berry et~al.}(1979)\citenamefont{Berry, Balazs, Tabor,
  and Voros}}]{berry1979quantum}
\bibinfo{author}{\bibfnamefont{M.~V.} \bibnamefont{Berry}},
  \bibinfo{author}{\bibfnamefont{N.~L.} \bibnamefont{Balazs}},
  \bibinfo{author}{\bibfnamefont{M.}~\bibnamefont{Tabor}}, \bibnamefont{and}
  \bibinfo{author}{\bibfnamefont{A.}~\bibnamefont{Voros}},
  \bibinfo{journal}{Annals of Physics} \textbf{\bibinfo{volume}{122}},
  \bibinfo{pages}{26} (\bibinfo{year}{1979}).

\bibitem[{\citenamefont{Hannay and Berry}(1980)}]{Hannay1980}
\bibinfo{author}{\bibfnamefont{J.~H.} \bibnamefont{Hannay}} \bibnamefont{and}
  \bibinfo{author}{\bibfnamefont{M.~V.} \bibnamefont{Berry}},
  \bibinfo{journal}{Physica D} \textbf{\bibinfo{volume}{1}},
  \bibinfo{pages}{267} (\bibinfo{year}{1980}).

\bibitem[{\citenamefont{Chirikov and Shepelyansky}(2008)}]{DimaScholar}
\bibinfo{author}{\bibfnamefont{B.}~\bibnamefont{Chirikov}} \bibnamefont{and}
  \bibinfo{author}{\bibfnamefont{D.~L.} \bibnamefont{Shepelyansky}},
  \bibinfo{journal}{Scholarpedia} \textbf{\bibinfo{volume}{3}},
  \bibinfo{pages}{3350} (\bibinfo{year}{2008}).

\bibitem[{\citenamefont{Goussev et~al.}(2012)\citenamefont{Goussev, Jalabert,
  Pastawski, and Wisniacki}}]{DiegoScholar}
\bibinfo{author}{\bibfnamefont{A.}~\bibnamefont{Goussev}},
  \bibinfo{author}{\bibfnamefont{R.}~\bibnamefont{Jalabert}},
  \bibinfo{author}{\bibfnamefont{H.~M.} \bibnamefont{Pastawski}},
  \bibnamefont{and} \bibinfo{author}{\bibfnamefont{D.~A.}
  \bibnamefont{Wisniacki}}, \bibinfo{journal}{Scholarpedia}
  \textbf{\bibinfo{volume}{7}}, \bibinfo{pages}{11687} (\bibinfo{year}{2012}).

\bibitem[{\citenamefont{Schack}(1998)}]{Schack1998}
\bibinfo{author}{\bibfnamefont{R.}~\bibnamefont{Schack}},
  \bibinfo{journal}{Phys. Rev. A} \textbf{\bibinfo{volume}{57}},
  \bibinfo{pages}{1634} (\bibinfo{year}{1998}).

\bibitem[{\citenamefont{Georgeot and Shepelyansky}(2001)}]{Bertrand2001}
\bibinfo{author}{\bibfnamefont{B.}~\bibnamefont{Georgeot}} \bibnamefont{and}
  \bibinfo{author}{\bibfnamefont{D.~L.} \bibnamefont{Shepelyansky}},
  \bibinfo{journal}{Phys. Rev. Lett.} \textbf{\bibinfo{volume}{86}},
  \bibinfo{pages}{5393} (\bibinfo{year}{2001}).

\bibitem[{\citenamefont{Weinstein et~al.}(2002)\citenamefont{Weinstein, Lloyd,
  Emerson, and Cory}}]{Weinstein2002}
\bibinfo{author}{\bibfnamefont{Y.}~\bibnamefont{Weinstein}},
  \bibinfo{author}{\bibfnamefont{S.}~\bibnamefont{Lloyd}},
  \bibinfo{author}{\bibfnamefont{J.}~\bibnamefont{Emerson}}, \bibnamefont{and}
  \bibinfo{author}{\bibfnamefont{D.}~\bibnamefont{Cory}},
  \bibinfo{journal}{Phys. Rev. Lett.} \textbf{\bibinfo{volume}{89}},
  \bibinfo{pages}{284102}
  (\bibinfo{year}{2002}).

\bibitem[{\citenamefont{L\'evi and Georgeot}(2004)}]{Bertrand2004}
\bibinfo{author}{\bibfnamefont{B.}~\bibnamefont{L\'evi}} \bibnamefont{and}
  \bibinfo{author}{\bibfnamefont{B.}~\bibnamefont{Georgeot}},
  \bibinfo{journal}{Phys. Rev. E} \textbf{\bibinfo{volume}{70}},
  \bibinfo{pages}{056218} (\bibinfo{year}{2004}).

\bibitem[{\citenamefont{Schwinger}(1960)}]{schwinger}
\bibinfo{author}{\bibfnamefont{J.}~\bibnamefont{Schwinger}},
  \bibinfo{journal}{Proc. Natl. Acad. Sci. U. S. A.} \textbf{\bibinfo{volume}{46}},
  \bibinfo{pages}{570} (\bibinfo{year}{1960}).

\bibitem[{\citenamefont{Ketzmerick et~al.}(2003)\citenamefont{Ketzmerick,
  Kruse, and Geisel}}]{geisel}
\bibinfo{author}{\bibfnamefont{R.}~\bibnamefont{Ketzmerick}},
  \bibinfo{author}{\bibfnamefont{K.}~\bibnamefont{Kruse}}, \bibnamefont{and}
  \bibinfo{author}{\bibfnamefont{T.}~\bibnamefont{Geisel}},
  \bibinfo{journal}{Physica D} \textbf{\bibinfo{volume}{131}},
  \bibinfo{pages}{247} (\bibinfo{year}{2003}).

\bibitem[{\citenamefont{Lichtenberg and Lieberman}(1983)}]{Licht}
\bibinfo{author}{\bibfnamefont{A.~J.} \bibnamefont{Lichtenberg}}
  \bibnamefont{and} \bibinfo{author}{\bibfnamefont{M.~A.}
  \bibnamefont{Lieberman}}, \emph{\bibinfo{title}{Regular and Stochastic
  Motion}}, Applied Mathematical Sciences 38 (\bibinfo{publisher}{Springer, New
  York}, \bibinfo{year}{1983}), ISBN \bibinfo{isbn}{9780387977454}.

\bibitem[{\citenamefont{Greene}(1979)}]{Greene1979}
\bibinfo{author}{\bibfnamefont{J.~M.} \bibnamefont{Greene}},
  \bibinfo{journal}{J. Math. Phys.} \textbf{\bibinfo{volume}{20}},
  \bibinfo{pages}{1183} (\bibinfo{year}{1979}).

\bibitem[{\citenamefont{Artuso}(2011)}]{ArtusoScholar}
\bibinfo{author}{\bibfnamefont{R.}~\bibnamefont{Artuso}},
  \bibinfo{journal}{Scholarpedia} \textbf{\bibinfo{volume}{6}},
  \bibinfo{pages}{10462} (\bibinfo{year}{2011}).

\bibitem[{\citenamefont{Leboeuf et~al.}(1990)\citenamefont{Leboeuf, Kurchan,
  Feingold, and Arovas}}]{leboeuf1990}
\bibinfo{author}{\bibfnamefont{P.}~\bibnamefont{Leboeuf}},
  \bibinfo{author}{\bibfnamefont{J.}~\bibnamefont{Kurchan}},
  \bibinfo{author}{\bibfnamefont{M.}~\bibnamefont{Feingold}}, \bibnamefont{and}
  \bibinfo{author}{\bibfnamefont{D.}~\bibnamefont{Arovas}},
  \bibinfo{journal}{Phys. Rev. Lett.} \textbf{\bibinfo{volume}{65}},
  \bibinfo{pages}{3076} (\bibinfo{year}{1990}).

\bibitem[{\citenamefont{Borgonovi and
  Shepelyansky}(1995)}]{borgonovi1995spectral}
\bibinfo{author}{\bibfnamefont{F.}~\bibnamefont{Borgonovi}} \bibnamefont{and}
  \bibinfo{author}{\bibfnamefont{D.}~\bibnamefont{Shepelyansky}},
  \bibinfo{journal}{EPL (Europhysics Letters)} \textbf{\bibinfo{volume}{29}},
  \bibinfo{pages}{117} (\bibinfo{year}{1995}).

\bibitem[{\citenamefont{Steinigeweg et~al.}(2013)\citenamefont{Steinigeweg,
  Herbrych, and Prelov\ifmmode~\check{s}\else
  \v{s}\fi{}ek}}]{PhysRevE.87.012118}
\bibinfo{author}{\bibfnamefont{R.}~\bibnamefont{Steinigeweg}},
  \bibinfo{author}{\bibfnamefont{J.}~\bibnamefont{Herbrych}}, \bibnamefont{and}
  \bibinfo{author}{\bibfnamefont{P.}~\bibnamefont{Prelov\ifmmode~\check{s}\else
  \v{s}\fi{}ek}}, \bibinfo{journal}{Phys. Rev. E}
  \textbf{\bibinfo{volume}{87}}, \bibinfo{pages}{012118}
  (\bibinfo{year}{2013}).

\bibitem[{\citenamefont{Gogolin et~al.}(2011)\citenamefont{Gogolin, M\"uller,
  and Eisert}}]{PhysRevLett.106.040401}
\bibinfo{author}{\bibfnamefont{C.}~\bibnamefont{Gogolin}},
  \bibinfo{author}{\bibfnamefont{M.~P.} \bibnamefont{M\"uller}},
  \bibnamefont{and} \bibinfo{author}{\bibfnamefont{J.}~\bibnamefont{Eisert}},
  \bibinfo{journal}{Phys. Rev. Lett.} \textbf{\bibinfo{volume}{106}},
  \bibinfo{pages}{040401} (\bibinfo{year}{2011}).

\bibitem[{\citenamefont{Marcuzzi et~al.}(2013)\citenamefont{Marcuzzi, Marino,
  Gambassi, and Silva}}]{PhysRevLett.111.197203}
\bibinfo{author}{\bibfnamefont{M.}~\bibnamefont{Marcuzzi}},
  \bibinfo{author}{\bibfnamefont{J.}~\bibnamefont{Marino}},
  \bibinfo{author}{\bibfnamefont{A.}~\bibnamefont{Gambassi}}, \bibnamefont{and}
  \bibinfo{author}{\bibfnamefont{A.}~\bibnamefont{Silva}},
  \bibinfo{journal}{Phys. Rev. Lett.} \textbf{\bibinfo{volume}{111}},
  \bibinfo{pages}{197203} (\bibinfo{year}{2013}).

\bibitem[{\citenamefont{Alba}(2015)}]{PhysRevB.91.155123}
\bibinfo{author}{\bibfnamefont{V.}~\bibnamefont{Alba}}, \bibinfo{journal}{Phys.
  Rev. B} \textbf{\bibinfo{volume}{91}}, \bibinfo{pages}{155123}
  (\bibinfo{year}{2015}).

\bibitem[{\citenamefont{Imbrie}(2016{\natexlab{a}})}]{Imbrie2016}
\bibinfo{author}{\bibfnamefont{J.~Z.} \bibnamefont{Imbrie}},
  \bibinfo{journal}{J. Stat. Phys.} \textbf{\bibinfo{volume}{163}},
  \bibinfo{pages}{998} (\bibinfo{year}{2016}{\natexlab{a}}).

\bibitem[{\citenamefont{Bardarson et~al.}(2012)\citenamefont{Bardarson,
  Pollmann, and Moore}}]{PhysRevLett.109.017202}
\bibinfo{author}{\bibfnamefont{J.~H.} \bibnamefont{Bardarson}},
  \bibinfo{author}{\bibfnamefont{F.}~\bibnamefont{Pollmann}}, \bibnamefont{and}
  \bibinfo{author}{\bibfnamefont{J.~E.} \bibnamefont{Moore}},
  \bibinfo{journal}{Phys. Rev. Lett.} \textbf{\bibinfo{volume}{109}},
  \bibinfo{pages}{017202} (\bibinfo{year}{2012}).

\bibitem[{\citenamefont{Imbrie}(2016{\natexlab{b}})}]{PhysRevLett.117.027201}
\bibinfo{author}{\bibfnamefont{J.~Z.} \bibnamefont{Imbrie}},
  \bibinfo{journal}{Phys. Rev. Lett.} \textbf{\bibinfo{volume}{117}},
  \bibinfo{pages}{027201} (\bibinfo{year}{2016}{\natexlab{b}}).

\bibitem[{\citenamefont{F~Santos et~al.}(2012)\citenamefont{F~Santos,
  Borgonovi, and Izrailev}}]{Borgonovi}
\bibinfo{author}{\bibfnamefont{L.}~\bibnamefont{F~Santos}},
  \bibinfo{author}{\bibfnamefont{F.}~\bibnamefont{Borgonovi}},
  \bibnamefont{and} \bibinfo{author}{\bibfnamefont{F.}~\bibnamefont{Izrailev}},
  \bibinfo{journal}{Phys. Rev. E} \textbf{\bibinfo{volume}{85}},
  \bibinfo{pages}{036209} (\bibinfo{year}{2012}).

\bibitem[{\citenamefont{Sachdev}(2001)}]{sachdev2001quantum}
\bibinfo{author}{\bibfnamefont{S.}~\bibnamefont{Sachdev}},
  \emph{\bibinfo{title}{Quantum Phase Transitions}}
  (\bibinfo{publisher}{Cambridge University Press, Cambridge}, \bibinfo{year}{2001}), ISBN
  \bibinfo{isbn}{9780521004541}.

\bibitem[{\citenamefont{Karthik et~al.}(2007)\citenamefont{Karthik, Sharma, and
  Lakshminarayan}}]{tilt}
\bibinfo{author}{\bibfnamefont{J.}~\bibnamefont{Karthik}},
  \bibinfo{author}{\bibfnamefont{A.}~\bibnamefont{Sharma}}, \bibnamefont{and}
  \bibinfo{author}{\bibfnamefont{A.}~\bibnamefont{Lakshminarayan}},
  \bibinfo{journal}{Phys. Rev. A} \textbf{\bibinfo{volume}{75}},
  \bibinfo{pages}{022304} (\bibinfo{year}{2007}).

\bibitem[{\citenamefont{Lin and Motrunich}(2018)}]{Rusos}
\bibinfo{author}{\bibfnamefont{C.-J.} \bibnamefont{Lin}} \bibnamefont{and}
  \bibinfo{author}{\bibfnamefont{O.~I.} \bibnamefont{Motrunich}},
  \bibinfo{journal}{Phys. Rev. B} \textbf{\bibinfo{volume}{97}},
  \bibinfo{pages}{144304} (\bibinfo{year}{2018}).

\bibitem[{\citenamefont{\ifmmode \check{Z}\else
  \v{Z}\fi{}nidari\ifmmode~\check{c}\else \v{c}\fi{}
  et~al.}(2008)\citenamefont{\ifmmode \check{Z}\else
  \v{Z}\fi{}nidari\ifmmode~\check{c}\else \v{c}\fi{}, Prosen, and
  Prelov\ifmmode~\check{s}\else \v{s}\fi{}ek}}]{Marko2008}
\bibinfo{author}{\bibfnamefont{M.}~\bibnamefont{\ifmmode \check{Z}\else
  \v{Z}\fi{}nidari\ifmmode~\check{c}\else \v{c}\fi{}}},
  \bibinfo{author}{\bibfnamefont{T.~c.~v.} \bibnamefont{Prosen}},
  \bibnamefont{and}
  \bibinfo{author}{\bibfnamefont{P.}~\bibnamefont{Prelov\ifmmode~\check{s}\else
  \v{s}\fi{}ek}}, \bibinfo{journal}{Phys. Rev. B}
  \textbf{\bibinfo{volume}{77}}, \bibinfo{pages}{064426}
  (\bibinfo{year}{2008}).

\bibitem[{\citenamefont{Pal and Huse}(2010)}]{Pal2010}
\bibinfo{author}{\bibfnamefont{A.}~\bibnamefont{Pal}} \bibnamefont{and}
  \bibinfo{author}{\bibfnamefont{D.~A.} \bibnamefont{Huse}},
  \bibinfo{journal}{Phys. Rev. B} \textbf{\bibinfo{volume}{82}},
  \bibinfo{pages}{174411} (\bibinfo{year}{2010}).

\bibitem[{\citenamefont{De~Luca and Scardicchio}(2013)}]{de2013ergodicity}
\bibinfo{author}{\bibfnamefont{A.}~\bibnamefont{De~Luca}} \bibnamefont{and}
  \bibinfo{author}{\bibfnamefont{A.}~\bibnamefont{Scardicchio}},
  \bibinfo{journal}{EPL (Europhysics Letters)} \textbf{\bibinfo{volume}{101}},
  \bibinfo{pages}{37003} (\bibinfo{year}{2013}).

\bibitem[{\citenamefont{Bauer and Nayak}(2013)}]{bauer2013area}
\bibinfo{author}{\bibfnamefont{B.}~\bibnamefont{Bauer}} \bibnamefont{and}
  \bibinfo{author}{\bibfnamefont{C.}~\bibnamefont{Nayak}}, \bibinfo{journal}{J.
  Stat. Mech.: Theo. Exp.} \textbf{\bibinfo{volume}{2013}},
  \bibinfo{pages}{P09005} (\bibinfo{year}{2013}).

\bibitem[{\citenamefont{Luitz et~al.}(2015)\citenamefont{Luitz, Laflorencie,
  and Alet}}]{Luitz}
\bibinfo{author}{\bibfnamefont{D.~J.} \bibnamefont{Luitz}},
  \bibinfo{author}{\bibfnamefont{N.}~\bibnamefont{Laflorencie}},
  \bibnamefont{and} \bibinfo{author}{\bibfnamefont{F.}~\bibnamefont{Alet}},
  \bibinfo{journal}{Phys. Rev. B} \textbf{\bibinfo{volume}{91}},
  \bibinfo{pages}{081103} (\bibinfo{year}{2015}).

\bibitem[{\citenamefont{Avishai et~al.}(2002)\citenamefont{Avishai, Richert,
  and Berkovits}}]{Avishai2002}
\bibinfo{author}{\bibfnamefont{Y.}~\bibnamefont{Avishai}},
  \bibinfo{author}{\bibfnamefont{J.}~\bibnamefont{Richert}}, \bibnamefont{and}
  \bibinfo{author}{\bibfnamefont{R.}~\bibnamefont{Berkovits}},
  \bibinfo{journal}{Phys. Rev. B} \textbf{\bibinfo{volume}{66}},
  \bibinfo{pages}{052416} (\bibinfo{year}{2002}).

\bibitem[{\citenamefont{Santos}(2004)}]{santos2004integrability}
\bibinfo{author}{\bibfnamefont{L.}~\bibnamefont{Santos}}, \bibinfo{journal}{J.
  Phys. A: Math. Gen.} \textbf{\bibinfo{volume}{37}}, \bibinfo{pages}{4723}
  (\bibinfo{year}{2004}).

\bibitem[{\citenamefont{Santos et~al.}(2005)\citenamefont{Santos, Dykman,
  Shapiro, and Izrailev}}]{Santos2005}
\bibinfo{author}{\bibfnamefont{L.~F.} \bibnamefont{Santos}},
  \bibinfo{author}{\bibfnamefont{M.~I.} \bibnamefont{Dykman}},
  \bibinfo{author}{\bibfnamefont{M.}~\bibnamefont{Shapiro}}, \bibnamefont{and}
  \bibinfo{author}{\bibfnamefont{F.~M.} \bibnamefont{Izrailev}},
  \bibinfo{journal}{Phys. Rev. A} \textbf{\bibinfo{volume}{71}},
  \bibinfo{pages}{012317} (\bibinfo{year}{2005}).

\bibitem[{\citenamefont{Alet and Laflorencie}(2018)}]{ALET2018}
\bibinfo{author}{\bibfnamefont{F.}~\bibnamefont{Alet}} \bibnamefont{and}
  \bibinfo{author}{\bibfnamefont{N.}~\bibnamefont{Laflorencie}},
  \bibinfo{journal}{Comptes Rendus Physique} \textbf{\bibinfo{volume}{19}},
  \bibinfo{pages}{498} (\bibinfo{year}{2018}).

\bibitem[{\citenamefont{Dukesz et~al.}(2009)\citenamefont{Dukesz, Zilbergerts,
  and Santos}}]{Dukesz_2009}
\bibinfo{author}{\bibfnamefont{F.}~\bibnamefont{Dukesz}},
  \bibinfo{author}{\bibfnamefont{M.}~\bibnamefont{Zilbergerts}},
  \bibnamefont{and} \bibinfo{author}{\bibfnamefont{L.~F.}
  \bibnamefont{Santos}}, \bibinfo{journal}{New J. Phys}
  \textbf{\bibinfo{volume}{11}}, \bibinfo{pages}{043026}
  (\bibinfo{year}{2009}).

\bibitem[{\citenamefont{Oteo}(1991)}]{Oteo:1991kg}
\bibinfo{author}{\bibfnamefont{J.~A.} \bibnamefont{Oteo}}, \bibinfo{journal}{J.
  Math. Phys.} \textbf{\bibinfo{volume}{32}}, \bibinfo{pages}{419}
  (\bibinfo{year}{1991}).

\end{thebibliography}
\providecommand{\noopsort}[1]{}\providecommand{\singleletter}[1]{#1}%

\end{document}
%